\documentclass[lettersize,journal]{IEEEtran}
\IEEEoverridecommandlockouts
\ifCLASSINFOpdf
\else
\fi
\usepackage{graphicx}
\DeclareUnicodeCharacter{202F}{\,}
\usepackage{xcolor}
\usepackage{hhline}
\usepackage{amsmath,amsfonts}
\usepackage{cleveref}
\Crefformat{figure}{#2Fig.~#1#3}
\Crefmultiformat{figure}{Figs.~#2#1#3}{ and~#2#1#3}{, #2#1#3}{ and~#2#1#3}
\usepackage{cuted}
\usepackage{tipa}
\usepackage{tipx}
\usepackage{scalerel}
\usepackage{textcomp}
\usepackage{threeparttable}
\usepackage{tablefootnote}
\usepackage{algpseudocode}
\usepackage{algorithm}
\usepackage{url}
\usepackage{multirow}
\usepackage{multicol}
\usepackage{soul}
\usepackage[normalem]{ulem}
\usepackage{comment}
\usepackage{diagbox}
\usepackage{amssymb}

\usepackage{array}
\usepackage[caption=false,font=normalsize,labelfont=sf,textfont=sf]{subfig}
\usepackage{stfloats}
\usepackage{verbatim}
\usepackage{multirow}
\usepackage{adjustbox}
\usepackage[utf8]{inputenc}
\usepackage[mathscr]{euscript}
\usepackage{caption}
\usepackage{subcaption}
\usepackage{bm}
\usepackage{comment}

\usepackage[backend=biber,,citestyle=numeric-comp,bibstyle=ieee,sorting=none,minbibnames=1,maxbibnames=6,doi=true,url=false,eprint=false]{biblatex}
\usepackage{acronym}

\def\BibTeX{{\rm B\kern-.05em{\sc i\kern-.025em b}\kern-.08em
    T\kern-.1667em\lower.7ex\hbox{E}\kern-.125emX}}
\usepackage{balance}

\begin{document}

\newacro{VLS}{voltage level shifter}
\newacro{HVLS}{hybrid \ac{VLS}}
\newacro{ADC}{analog-to-digital converter}
\newacro{DAC}{digital-to-analog converter}
\newacro{PLL}{phase-locked loop}
\newacro{LDO}{low-dropout regulator}
\newacro{CMLS}{current-mode level shifter}
\newacro{DDRF}{direct digital-to-RF}
\newacro{UBB}{ultra-broadband}
\newacro{RAN}{radio access network}
\newacro{SCPA}{switched-capacitor power amplifier}
\newacro{DTC}{digital-to-time converter}
\newacro{RF-DAC}{radio-frequency digital-to-analog converter}
\newacro{RF-ADC}{radio-frequency analog-to-digital converter}
\newacro{IF}{intermediate frequency}
\newacro{TTD}{true-time-delay}
\newacro{DDMW}{direct digital-to-mm-wave}
\newacro{TX}{transmitter}
\newacro{RX}{receiver}
\newacro{TRX}{transceiver}
\newacro{SNR}{signal-to-noise ratio}
\newacro{BS}{basestation}
\newacro{UE}{user equipment}
\newacro{AF}{array factor}
\newacro{TIPA}{two-input power amplifier}
\newacro{RF}{radio frequency}
\newacro{RAN}{radio access network}
\newacro{QoS}{quality of service}
\newacro{V2V}{vehicle-to-vehicle}
\newacro{OOB}{out-of-band}
\newacro{TAF}{time-approximation filter}
\newacro{LUT}{look-up table}
\newacro{OSR}{over-sampling rate}
\newacro{BER}{bit error rate}
\newacro{PAA}{phased-array antenna}
\newacro{AoA}{angle of arrival}
\newacro{AoD}{angle of departure}
\newacro{ML}{machine learning}
\newacro{AI}{artificial intelligence}
\newacro{FoM}{figure of merit}
\newacro{TI}{time-interleaved}
\newacro{NN}{neural network}
\newacro{CNN}{convolutional \ac{NN}}
\newacro{LoS}{line-of-sight}
\newacro{PVT}{process, voltage, and temperature}
\newacro{AR}{augmented reality}
\newacro{FDD}{frequency division duplexing}
\newacro{DNN}{deep \ac{NN}}
\newacro{LSTM}{long short-term memory}
\newacro{OAM}{orbital angular momentum}
\newacro{WDM}{wavelength division multiplexing}
\newacro{IoT}{internet-of-things}
\newacro{CMOS}{complementary metal-oxide-semiconductor}
\newacro{EDA}{electronic design automation}
\newacro{ATE}{automated test equipment}
\newacro{SDR}{software defined radio}
\newacro{FPGA}{field programmable gate array}
\newacro{LNA}{low noise amplifier}
\newacro{PA}{power amplifier}
\newacro{RFPA}{\ac{RF} \ac{PA}}
\newacro{PCM}{phase-change material}
\newacro{CR}{cognitive radio}
\newacro{SoC}{system-on-chip}
\newacro{RF-SoC}{radio-frequency system-on-chip}
\newacro{VT-ARC}{Virginia Tech Advanced Research Computing}
\newacro{GPU}{graphics processing unit}
\newacro{DSP}{digital signal processor}
\newacro{ASIC}{application specific IC}
\newacro{RC}{reservoir computing}
\newacro{FEM}{front-end module}
\newacro{MICS}{multi-function integrated circuits and systems}
\newacro{MIMO}{multiple-input, multiple-output}
\newacro{dMIMO}{distributed \ac{MIMO}}
\newacro{O-RAN}{open radio access network}
\newacro{O-RU}{open-radio unit}
\newacro{O-DU}{open-distributed unit}
\newacro{FR}{frequency-range}
\newacro{FR1}{\ac{FR} 1}
\newacro{FR3}{frequency-range 3}
\newacro{RRH}{remote-radio head}
\newacro{CIM}{compute-in-memory}
\newacro{COTS}{commercial, off-the-shelf}
\newacro{CDR}{clock and data recovery}
\newacro{LOS}{line of sight}
\newacro{CCI}{Commonwealth Cyber Initiative}
\newacro{PLL}{phase-locked loop}
\newacro{PDK}{process development kit}
\newacro{OTA}{over-the-air testing}
\newacro{FR2}{\ac{FR} 2}
\newacro{DFT}{discrete Fourier transform}
\newacro{SRAM}{Static-RAM}
\newacro{RIC}{Radio Intelligent Controller}
\newacro{near-RT}{near-real-time}
\newacro{IC}{integrated circuit}
\newacro{EIRP}{Effective Isotropic Radiated Power}
\newacro{NNBF}{Neural Network-Based Beamformer}
\newacro{BPA}{Beam Pointing Angle}
\newacro{mmWave}{millimeter-wave}
\newacro{QoE}{quality-of-experience}
\newacro{RU}{radio unit}
\newacro{PCB}{printed circuit board}
\newacro{MMIC}{monolithic microwave integrated circuit}
\newacro{MU-MIMO}{multi-user MIMO}
\newacro{LMBPA}{load-modulated, balanced power amplifier}
\newacro{TRL}{technology readiness level}
\newacro{MRL}{manufacturing readiness level}
\newacro{MVDR}{minimum variance distortionless response}
\newacro{SINR}{signal to interference and noise ratio}
\newacro{RRH}{remote radio head}
\newacro{CU}{central unit}
\newacro{DU}{distributed unit}
\newacro{gNB}{gNodeB}
\newacro{FM}{frequency modulation}
\newacro{PM}{phase modulation}
\newacro{OFDM}{orthogonal frequency division multiplexing}
\newacro{AM}{amplitude modulation}
\newacro{QAM}{quadrature \ac{AM}}
\newacro{PAE}{power added efficiency}
\newacro{DE}{$\eta$}
\newacro{PAPR}{peak to average power ratio}
\newacro{SC}{single-carrier}
\newacro{ISI}{inter-symbol interference}
\newacro{RRC}{root-raised cosine}
\newacro{PDF}{probability density function}
\newacro{CCDF}{complementary cumulative distribution function}
\newacro{CDF}{cumulative distribution function}
\newacro{PMEPR}{peak-to-mean envelope power ratio}
\newacro{DUT}{device under test}
\newacro{DPA}{digital power amplifier}
\newacro{DCVS}{differential-cascade voltage shifter}
\newacro{CMLS}{current-mirror-based level shifter}
\newacro{WCMLS}{Wilson current-mirror level shifter}
\newacro{HVLS}{hybrid voltage level shifter}
\newacro{RF-SoC}{\ac{RF}-\ac{SoC}}

\title{A High-Speed Differential Hybrid Voltage Level Shifter With Simultaneous Multi-Level Outputs}

\author{Behdad Jamadi, \textit{Student Member, IEEE}, Meysam Sohani Darban, \textit{Student Member, IEEE},\\ Jungmin Lee, \textit{Student Member, IEEE}, Jeffrey Sean Walling, \textit{Senior Member, IEEE}.
\vspace{-5mm}
\thanks{This work has been submitted to the IEEE for possible publication. Copyright may be transferred without notice, after which this version may no longer be accessible. This work was supported by the National Science Foundation (NSF) under grant \#2314813 and GlobalFoundries' University Partner Program. (Corresponding author: Behdad Jamadi, e-mail: behdadjamadi@vt.edu). The authors are with the Bradley Department of Electrical and Computer Engineering, Virginia Tech, Blacksburg, VA 24061 USA. Behdad Jamadi and Meysam Sohani Darban contributed equally to this work.}}

\maketitle

\begin{abstract}
This paper presents a low-jitter \ac{HVLS} architecture for high-speed mixed-signal and digital power-amplifier applications. The proposed design employs a regenerative cross-coupled feedback network to simultaneously generate two synchronized voltage-domain outputs (e.g., the nominal supply voltage ($V_{\mathrm{DDL}}$) and one at twice its value ($V_{\mathrm{DDH}} = 2V_{\mathrm{DDL}}$)). This enables direct drive of cascoded class-D power amplifiers without additional delay-calibration circuitry, amongst other mixed-signal applications. A prototype \ac{HVLS} circuit, together with an impedance-matching network and a pre-driver for high-speed off-chip characterization, was fabricated in a 22-nm FD-SOI process technology. The total die area, including all interface circuitry, is 477$\times$462~$\mu$m$^{2}$, while the active area of the \ac{HVLS} core is 3.26$\times$2~$\mu$m$^{2}$. Measured at 12.2~GHz, the circuit dissipates 427.8~$\mu$W and achieves a measured rms jitter below 43~fs (10~Hz--500~kHz), limited by the noise floor of the measurement setup.
\end{abstract}


\IEEEpeerreviewmaketitle
\section{Introduction}
\label{sec_int}

Multi-domain supply architectures are widely used in digital and mixed-signal integrated circuits to reduce dynamic power consumption. Interfacing between these domains requires \acfp{VLS}~\cite{Serneels2006,Moghe2011,Wooters2010,Luo2014}, which translate logic levels from a lower to a higher-voltage domain. Because an \ac{RF-SoC} may contain a large number of level shifters, their power consumption, propagation delay, and area are major design considerations~\cite{ref7,ref4,ref6}.

\begin{figure}[h]
\centering
\includegraphics[width=0.5\textwidth]{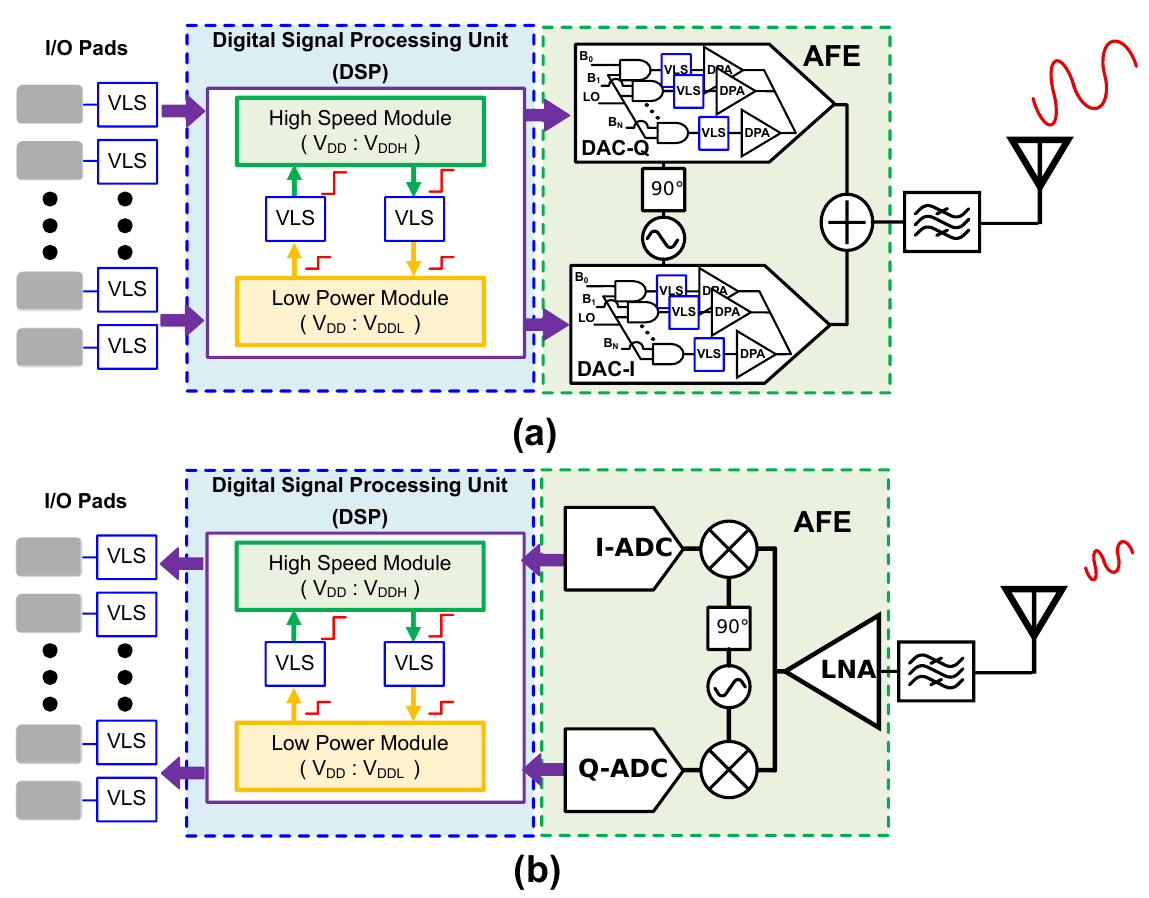}
\caption{RF \ac{TRX} architectures: a) conventional digital \ac{TX} with an \ac{RF-DAC} and digital back-end and b) conventional analog \ac{RX} with an \ac{RF-ADC} and digital back-end.}
\label{fig_ls_SoC}
\vspace{-2.5mm}
\end{figure}

With continued CMOS scaling, the speed gains of digital circuits are not matched in analog and RF circuits, whose intrinsic gain and output impedance degrade. Switched-mode architectures such as \acp{DPA}~\cite{Kavousian2008,Presti2009} and digital RF \acp{TRX}~\cite{Staszewski2005} therefore benefit directly from scaling and are replacing analog \ac{TRX} blocks in \acp{RF-SoC} (Fig.~\ref{fig_ls_SoC}).

One such example of an analog block designed solely from switching circuits is the \ac{SCPA} \cite{ref_scpa_2011, ref_tcasii_tx, ref_jsscc_scpa_2}. The output stage of the \ac{SCPA} is an inverter-based class-D \ac{PA} that faces challenges with increasing output power~\cite{ref_tcasii_dpa}. This limitation arises because class-D \acp{PA} use PMOS switches tied to the supply, which restricts the voltage swing, unlike inductor-biased topologies. To support larger output-voltage swings, \acp{SCPA} and related class-D PAs often use cascoded/stacked inverters, allowing operation at up to $2\times$ the nominal supply voltage when transistor gates are driven appropriately~\cite{ref_tcasii_dpa,ref_jsscc_scpa_2}. This approach enhances output power and efficiency, but requires a \ac{VLS} to drive the upper PMOS transistor. 

AC-coupled level shifting is an alternative way to drive the cascoded class-D stage. It is less attractive for \acp{SCPA} and digital \acp{TX}, however, because these architectures use many parallel cells, so the coupling-capacitor area scales with the cell count. Moreover, AC coupling does not remove the cell-to-cell timing errors caused by clock-distribution and routing mismatches across the larger array, which degrade the coherent combination and the output power~\cite{ref_scpa_routing}.

A key challenge in cascoded topologies is synchronizing the signals driving the PMOS and NMOS transistors. The PMOS typically switches between $\mathrm{V_{DD}}$ and $\mathrm{2V_{DD}}$, while the NMOS switches between $\mathrm{GND}$ and $\mathrm{V_{DD}}$. Previous solutions employed programmable delay chains~\cite{ref_tcasii_dpa,ref_jsscc_scpa_2}; however, these are sensitive to \ac{PVT} variations and require calibration, which increases power, area, and complexity. The lack of suitable high-speed \ac{VLS} designs has precluded practical implementation of a high-speed cascoded class-D topology.

To overcome these limitations, this paper introduces a low-jitter, power-efficient \acf{HVLS} employing several techniques: 1) positive feedback that introduces a negative-resistance
network with a small $|R_{eq}|$ (i.e., a large negative conductance),
which strengthens the regeneration and reduces jitter; 2) current steering branches in order to improve switching transition speed and reduce timing uncertainty, and 3) diode connected transistors to decrease the contention current, which makes the proposed \ac{HVLS} capable of generating both required voltage levels simultaneously while operating at $>12.2\rm ~GHz$. Its speed and efficiency make it suitable for emerging \ac{FR3}-band \acp{TRX} in next-generation wireless systems.

This paper is organized as follows. Section~\ref{sec_litrev} reviews and compares conventional \ac{VLS} topologies. Section~\ref{sec_analys} analyzes the impact of drive-signal timing mismatch on the output power of cascoded class-D \acp{PA}. Section~\ref{sec_prop} describes the proposed \ac{HVLS} design and its analytical model. Section~\ref{sec_sim} presents simulation results for both analog and digital applications, and Section~\ref{sec_res} presents measurement results, followed by Section~\ref{sec_conc}, which concludes the work.

\section{Review of Voltage Level-Shifters (VLS) }
\label{sec_litrev}

Most conventional \acp{VLS} belong to one of three families: \acp{DCVS}, \acp{CMLS}, and \acp{WCMLS}~\cite{ref2,ref3}, which trade off speed, power consumption, and conversion range differently.

\begin{figure*}[t]
\centering
\includegraphics[width=\textwidth]{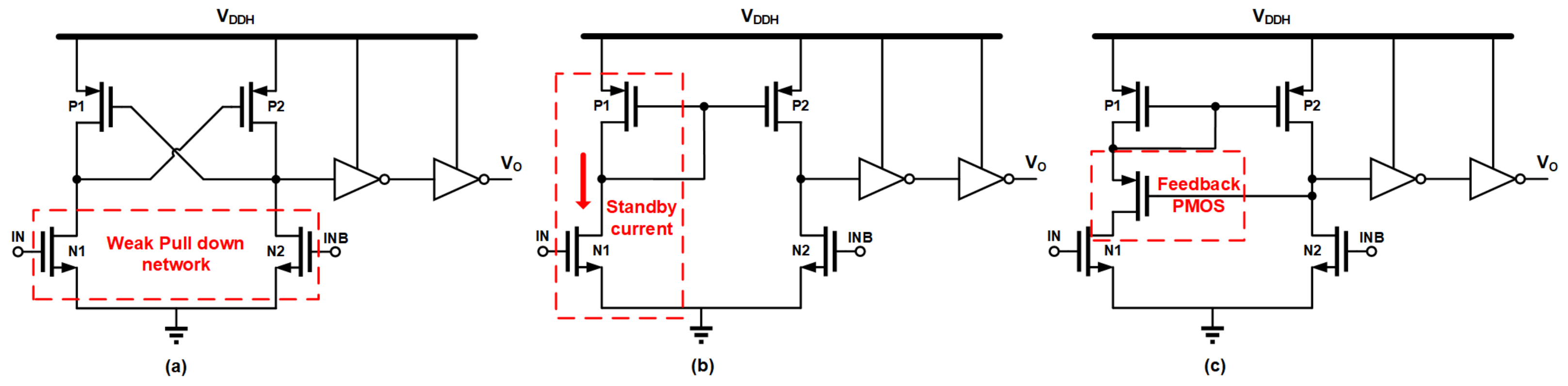}
\caption{Schematics of three types of \acp{VLS}: a) Type I: \ac{DCVS}, b) Type II: \ac{CMLS}, and c) Type III: \ac{WCMLS}.}
\label{fig_ls_litrev}
\vspace{-5mm}
\end{figure*}

The \ac{DCVS} topology in Fig.~\ref{fig_ls_litrev}(a) uses a cross-coupled PMOS pull-up network and a differential NMOS pull-down network~\cite{ref8}. Since the NMOS devices are driven by the lower supply ($V_{DDL}$) while the PMOS devices operate from $V_{DDH}$, the pull-up path can dominate, especially at low $V_{DDL}$, causing contention current, increased power, reduced speed, and limited conversion range. Mitigation requires either larger NMOS devices, with area/capacitance penalties, or a higher minimum $V_{DDL}$. The \ac{CMLS} in Fig.~\ref{fig_ls_litrev}(b) replaces the cross-coupled PMOS latch with a current mirror to reduce contention~\cite{ref6,ref2,ref3}; however, the diode-connected PMOS creates a continuous current path, resulting in high standby power. The \ac{WCMLS} in Fig.~\ref{fig_ls_litrev}(c) adds feedback PMOS to interrupt this static path~\cite{ref_wil,ref1}, reducing standby current but also limiting output-drive, increasing delay, degrading output swing, and potentially creating a floating feedback-gate node.

Several improvements have therefore been proposed. A current limiter in~\cite{ref_comp6} reduces \ac{DCVS} contention but also lowers output current and swing, while the limited-swing inverter in~\cite{ref_LSc} reduces power at the cost of incomplete pull-up transitions and additional static loss~\cite{ref_LSa}. For current-mirror-based designs, stacked multi-$V_T$ devices~\cite{ref_LSc} reduce contention but weaken the pull-up network; current-mismatch circuitry~\cite{ref_LSb} decouples signal paths but still suffers from contention during high-to-low transitions; and preamplifier-based designs~\cite{ref_LSd} improve robustness but increase transition time and area. Hybrid current-mirror/cross-coupled structures~\cite{ref_LSe,ref5} improve speed and operating frequency, although switching-induced contention remains.

These improvements do not remove the topology-dependent trade-off between operating frequency, delay, and power. To quantify it, the three baseline topologies were simulated at schematic level in 22-nm FD-SOI at the TT corner and 27$^{\circ}$C, with $V_{DDH}=0.9$~V (the nominal supply of the process), input rise/fall times of 1\% of the period, and a minimum-size inverter load ($\approx$0.48~fF), while $V_{DDL}$ was swept from 0.4~V (near $V_{th}$) to 0.8~V. All three topologies use identical device sizes in Table~\ref{table_size}, chosen for a common target of rail-to-rail output into this load at low power, because in an \ac{SCPA} the level shifter of every cell sees the same input driver and load.

\begin{figure}[t]
\centering
\includegraphics[width= 0.45\textwidth]{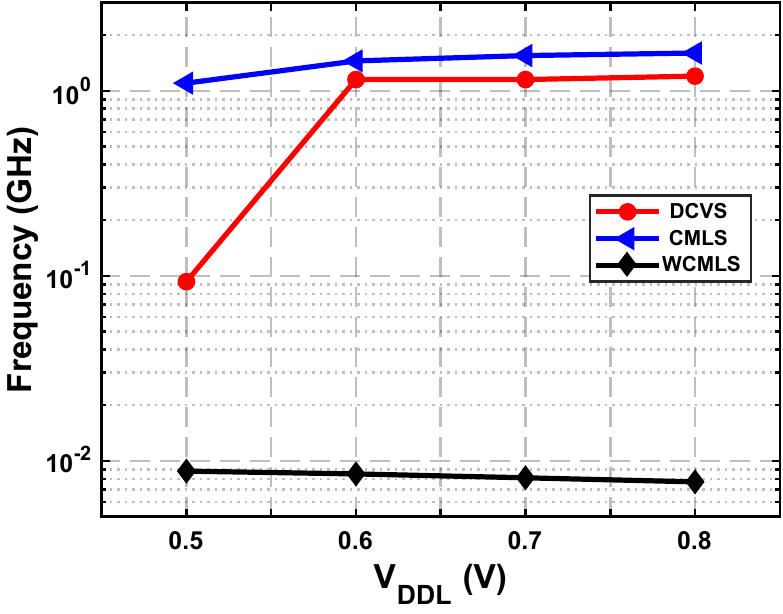}
\caption{Maximum operation frequency of the three types of \acp{VLS} by adjusting the input voltage ($V_{DDL}$).}
\label{fig_comp_fin}
\vspace{-5mm}
\end{figure}

\begin{table}[b]
\centering
\caption{Common \ac{VLS} Topology Devices' Size  (W/L)}
\resizebox{\columnwidth}{!}{%
\begin{tabular}{|c|c|c|c|}
\hline
\multicolumn{1}{|c|}{\textbf{Topology}} &
\multicolumn{1}{c|}{\ac{DCVS}} &
\multicolumn{1}{c|}{\ac{CMLS}} &
\multicolumn{1}{c|}{\ac{WCMLS}} \\ \hhline{|-|-|-|-|}
\textbf{N1,N2} & 400n / 20n & 400n / 20n & 400n / 20n \\ \hline
\textbf{P1,P2} & 250n / 20n & 250n / 20n & 250n / 20n \\ \hline
\textbf{Feedback PMOS} & - & - & 250n / 20n \\ \hline
\end{tabular}%
}
\label{table_size}
\end{table}

Fig.~\ref{fig_comp_fin} shows the maximum operating frequency versus $V_{DDL}$. The \ac{CMLS} and \ac{DCVS} reach comparable frequencies over most of the range, whereas the \ac{WCMLS} is limited to a few MHz: suppressing its static current also removes most of its switching current.

\begin{figure}[t]
\centering
\centerline{\includegraphics[width= 0.45\textwidth]{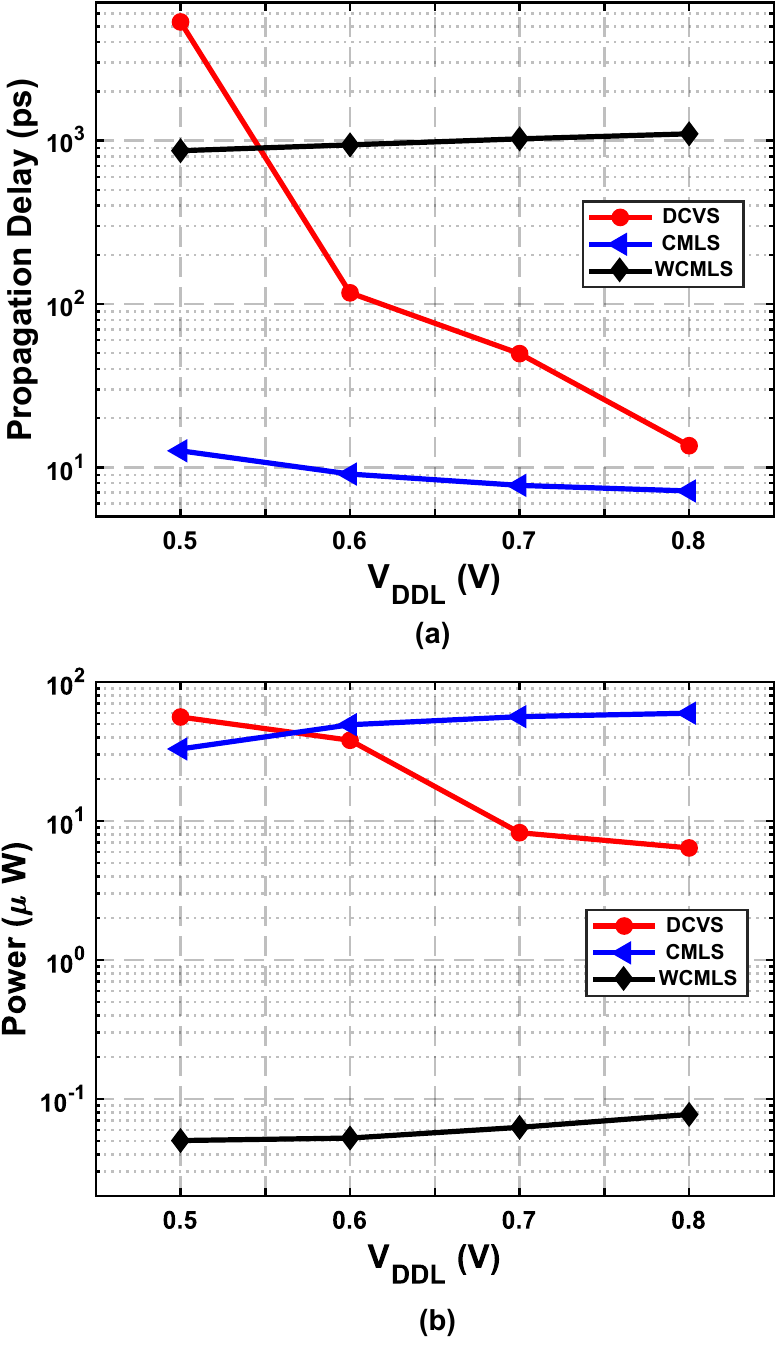}}
\caption{(a) Propagation delay and (b) total (static plus dynamic) power drawn from both supplies of the three types of \acp{VLS} versus the input voltage, each toggled at its maximum operating frequency from Fig.~\ref{fig_comp_fin}.}
\label{fig_comp_pd}
\vspace{-5mm}
\end{figure}

Fig.~\ref{fig_comp_pd} shows the corresponding propagation delay and power consumption. The \ac{CMLS} has the lowest delay but the highest power over most of the range, because of the static current in its mirror branch. The \ac{WCMLS} draws much less current, but its limited transition current makes its delay nearly an order of magnitude larger. The \ac{DCVS} depends strongly on $V_{DDL}$, since its NMOS pull-down current is set by the input voltage: as $V_{DDL}$ rises from 0.5~V to 0.8~V, its delay drops by nearly three orders of magnitude and, as contention falls, its power decreases (red lines in Fig.~\ref{fig_comp_pd}). Optimizing any one of these metrics is therefore insufficient for a level shifter intended for both low-voltage and high-speed operation.

Based on the above analysis, several key design requirements for an efficient voltage level shifter can be identified:
\begin{enumerate}
    \item Minimized contention current to ensure efficient voltage level conversion
    \item Fast switching under low input voltages
    \item Elimination of static current paths to reduce power consumption
    \item Compact area with minimal device count
    \item Scalable delay performance as the supply voltages converge.
\end{enumerate}

No existing \ac{VLS} architecture meets all of these requirements simultaneously, particularly when driving the stacked class-D \acp{PA} of \ac{SCPA} systems. The next section analyzes how drive-signal delay mismatch degrades such cascoded class-D \acp{PA}, which motivates the proposed level shifter.

\section{Effects of Delay Mismatch on Cascoded Class-D PA Output Power}
\label{sec_analys}

\begin{figure}[t]
\centering
\includegraphics[width=0.5\textwidth]{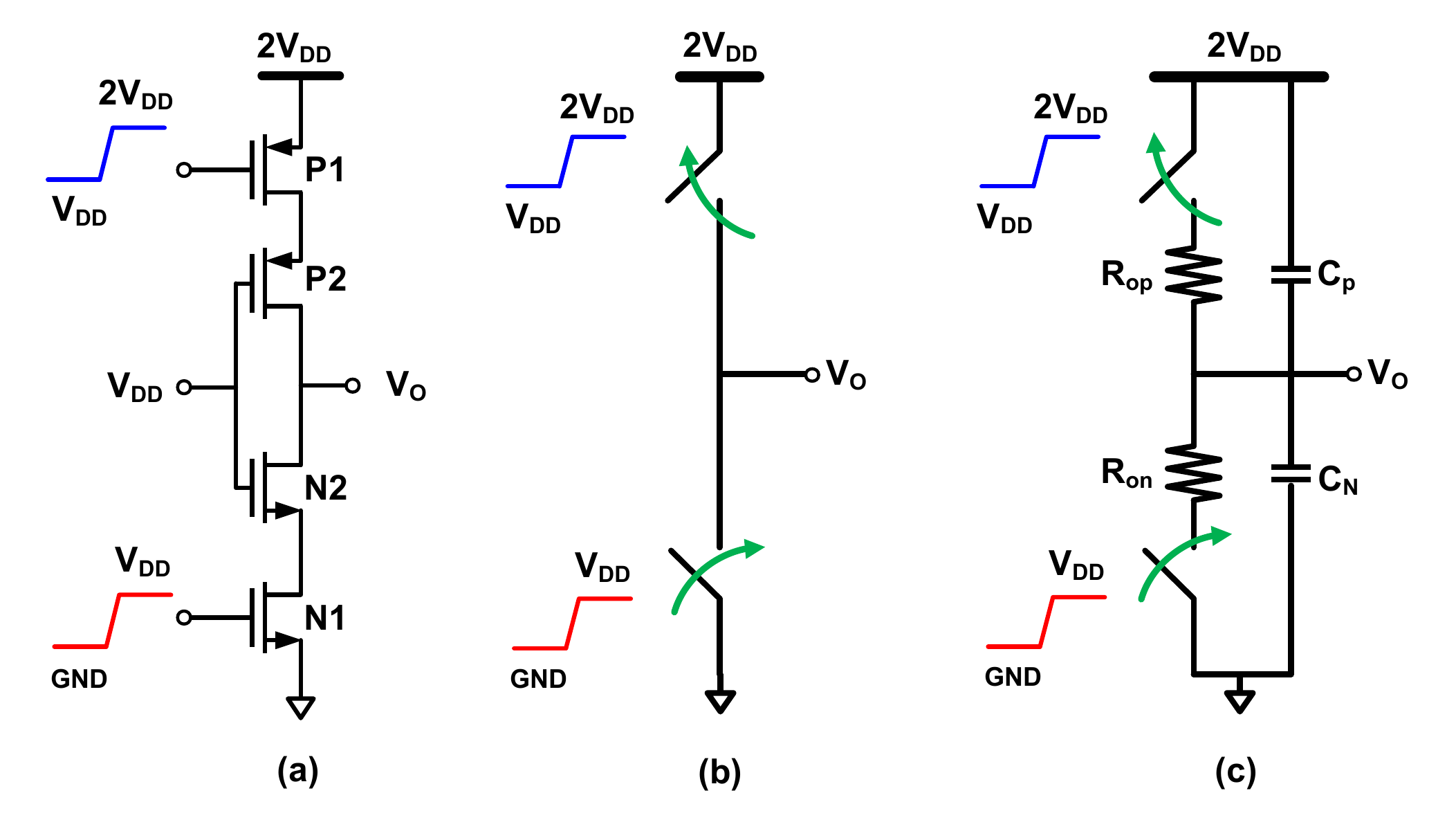}
\caption{Stacked class-D PA: a) schematic, b) ideal model, and c) non-ideal model.}
\label{fig_ls_sw}
\vspace{-5mm}
\end{figure}

\begin{figure}[t]
\centering
\includegraphics[width=0.5\textwidth]{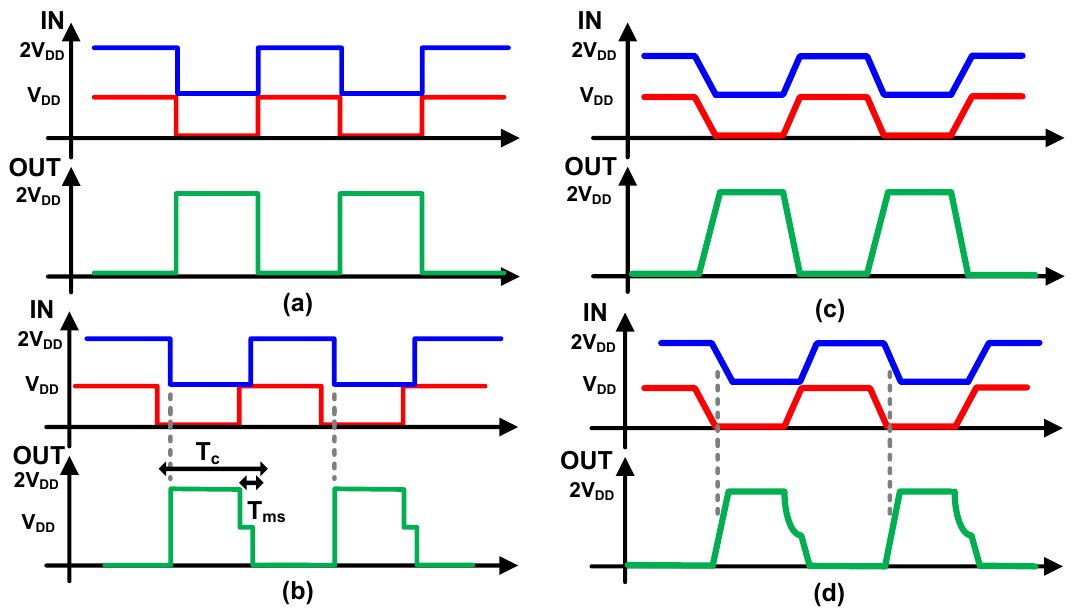}
\caption{Waveforms for cascoded class-D PA using an ideal switch model when driven by: a) ideally synchronous signals, and b) signals with delay mismatch. Using a non-ideal switch model when driven by: c) ideally synchronous signals, and d) signals with delay mismatch.}
\label{fig_ls_mm_wave_ideal}
\vspace{-5mm}
\end{figure}

Some applications require a level shifter to provide both the input-domain ($V_{DDL}$) and the output-domain ($V_{DDH}$) signals simultaneously, e.g., control signals that cross supply domains but must remain synchronous with their controller, or cascoded class-D \acp{PA}. In the cascoded switch of \Cref{fig_ls_sw}, the PMOS and NMOS devices connected to the supplies must be switched simultaneously by signals in two different voltage domains; any delay mismatch between them distorts the duty cycle and shortens the output pulse. Its impact is evaluated below for an ideal cascode class-D PA, with ideal switches, and a non-ideal one that includes the switch resistance and parasitic capacitance.

\subsection{Case Study I: Ideal Cascode Class-D PA}
Under ideal conditions, the switches have zero resistance, and the NMOS (PMOS) can be modeled as a pull-down (pull-up) device, as shown in \Cref{fig_ls_sw}~(b). For the ideal case where the signals are synchronized (\Cref{fig_ls_mm_wave_ideal}~(a)), the output power, $\mathrm{P_{out,i}}$, of the class-D PA with double the nominal supply voltage is calculated according to the first Fourier coefficient \cite{ref_scpa_2011}:

\begin{equation}
    P_{out,i} = \dfrac{8}{\pi^2}\cdot\dfrac{V_{DD}^2}{R_{opt}},
    \label{eq_poi}
\end{equation}

\noindent where $\mathrm{R_{opt}}$ is the optimum resistance seen from the matching network.

A delay mismatch in toggling the respective gate voltages of P1 and N1 results in a two-level step signal waveform, as shown in \Cref{fig_ls_mm_wave_ideal}~(b). The output signal has a voltage level of $\mathrm{2V_{DD}}$, with a pulse width of $\mathrm{T_{c} - T_{ms}}$ where $\mathrm{T_{c}}$ is $\dfrac{1}{ 2\cdot f_{c}}$, $f_{c}$ is the carrier frequency, and $\mathrm{T_{ms}}$ represents the time mismatch between signals driving PMOS and NMOS in cascode class-D PA. Hence, the output power of a stacked class-D PA under delay mismatch, $P_{out, ni}$, can be calculated as follows:
\begin{equation}
P_{out,ni} = P_{out,i}\cdot\left[\sin^{2}(\pi\cdot(\dfrac{1}{2} - \delta))+\dfrac{1}{4} \cdot \sin^{2}(\pi\delta)\right],
\label{eq_pni}
\end{equation}

\noindent where
\begin{equation}
    \delta  = \dfrac{T_{ms}}{2T_{c}}=\dfrac{T_{ms}}{T}.
\end{equation}

\subsection{Case Study II: Non-Ideal Cascode Class-D PA}
\begin{figure}[t]
\centering
\includegraphics[width=0.5\textwidth]{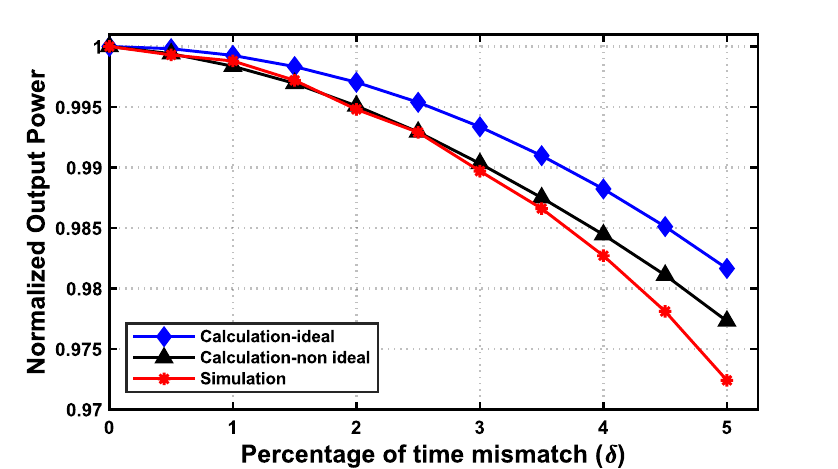}
\caption{Output power degradation of the stacked class-D PA in comparison with the percentage of the time mismatch  ($\delta$) between PMOS and NMOS driving signals.}
\label{fig_LS_pout_tmis}
\vspace{-5mm}
\end{figure}

With real transistors, the switch on-resistance and parasitic capacitance cannot be neglected (\Cref{fig_ls_sw}(c)), and the output no longer toggles abruptly between the rails. While only the PMOS devices conduct, they operate in triode with a small on-resistance ($\approx 0.05R_{opt}$), and the output power approximately follows~\eqref{eq_poi}. During the mismatch interval, however, both the NMOS and PMOS devices conduct in saturation, and the output time constant is $\tau=R_{out}C_{out}$, with:

\begin{equation}
    R_{out} = R_{on} || R_{op},
    \label{eq_rout}
\end{equation}
where $R_{op/on}$ is equivalent to:
\begin{equation}
    R_{op/on} = r_{o1,n/p} + r_{o2,n/p} + g_{m1,n/p} \cdot r_{o1,n/p} \cdot r_{o2,n/p}.
    \label{eq_ro}
\end{equation}

where $r_{oX,n/p}$ is the output resistance of each device during the transition, and
wherein both devices operate in the saturation region. The output capacitance of the cascode inverter is primarily determined by the intrinsic parasitic capacitance from the transistors.
\begin{equation}
    C_{out} = c_{gd,n} \cdot (1+Av^{-1}) + c_{gd,p} \cdot (1+Av^{-1}),
    \label{eq_cout}
\end{equation}
where $A_v$ is the intrinsic gain. Because $R_{out}\gg R_{opt}$ in this interval, the mismatch must be evaluated during the rise and fall times, and the output is modeled as a first-order RC charge/discharge while both devices are on.
Assuming that the delay mismatch is comparable to the rise and fall times but much smaller than the period, each rail-to-rail transition consists of a linear ramp, while the drive signals do not overlap, and an exponential segment, while they overlap and the pull-up and pull-down paths contend (\Cref{fig_ls_mm_wave_ideal}(d)). The fundamental Fourier coefficient is then the sum of those of the step, the ramp, $C_{rmp}$, and the exponential segment, $C_{exp}$:

\begin{equation}
    C_{rmp} = \dfrac{\Delta V}{\Delta t \cdot T} \cdot \dfrac{e^{-j\omega \Delta t}(j\omega\Delta t +1 ) - 1}{\omega^2},
    \label{eq_c1_ramp}
\end{equation}
and
\begin{equation}
    C_{exp} = \dfrac{\Delta V}{ T }\cdot [\dfrac{1-e^{-j\omega\Delta t}}{j\omega} - \dfrac{1-e^{-\Delta t (1/\tau + j\omega)}}{1/\tau + j\omega}],
    \label{eq_c1_exp}
\end{equation}
where $\omega=2\pi/T$, and $\Delta t$ and $\Delta V$ are the duration and voltage change of each segment.
\Cref{fig_LS_pout_tmis} compares the ideal model, the non-ideal model, and transistor-level simulation under nominal conditions at 27$^{\circ}$C. The output power decreases by about 3\% as the delay mismatch increases from 0 to 5\% of the carrier period; because a given timing mismatch is a larger fraction of a shorter period, the degradation grows with operating frequency. The non-ideal model agrees with the transistor-level results up to about 5\% mismatch and deviates gradually beyond it, because its piecewise ramp/exponential approximation omits higher-order nonlinear and parasitic effects. Over the practical range of mismatch, it predicts the degradation accurately and identifies its dominant mechanism.

The power loss grows with the overlap between the PMOS and NMOS drive signals. Because the loss is a fraction of the output power, it is the same for a single slice as for a complete \ac{SCPA} in which all slices share the same mismatch; in a large array, the cumulative cost of mismatch appears instead as shoot-through current during the overlap interval, which scales with the number of active slices and reduces the efficiency. In other mixed-signal applications, a lack of synchronization causes other undesired effects, such as increased harmonic content or loss of control stability. This motivates the proposed \ac{HVLS}, described in the next section, which provides a level-shifted output ($V_{DDH}$) and a synchronous output at the input logic levels ($V_{DDL}$).

\section{Proposed Hybrid Voltage Level Shifter}
\label{sec_prop}

\begin{figure}[t]
\centering
\includegraphics[width=0.5\textwidth]{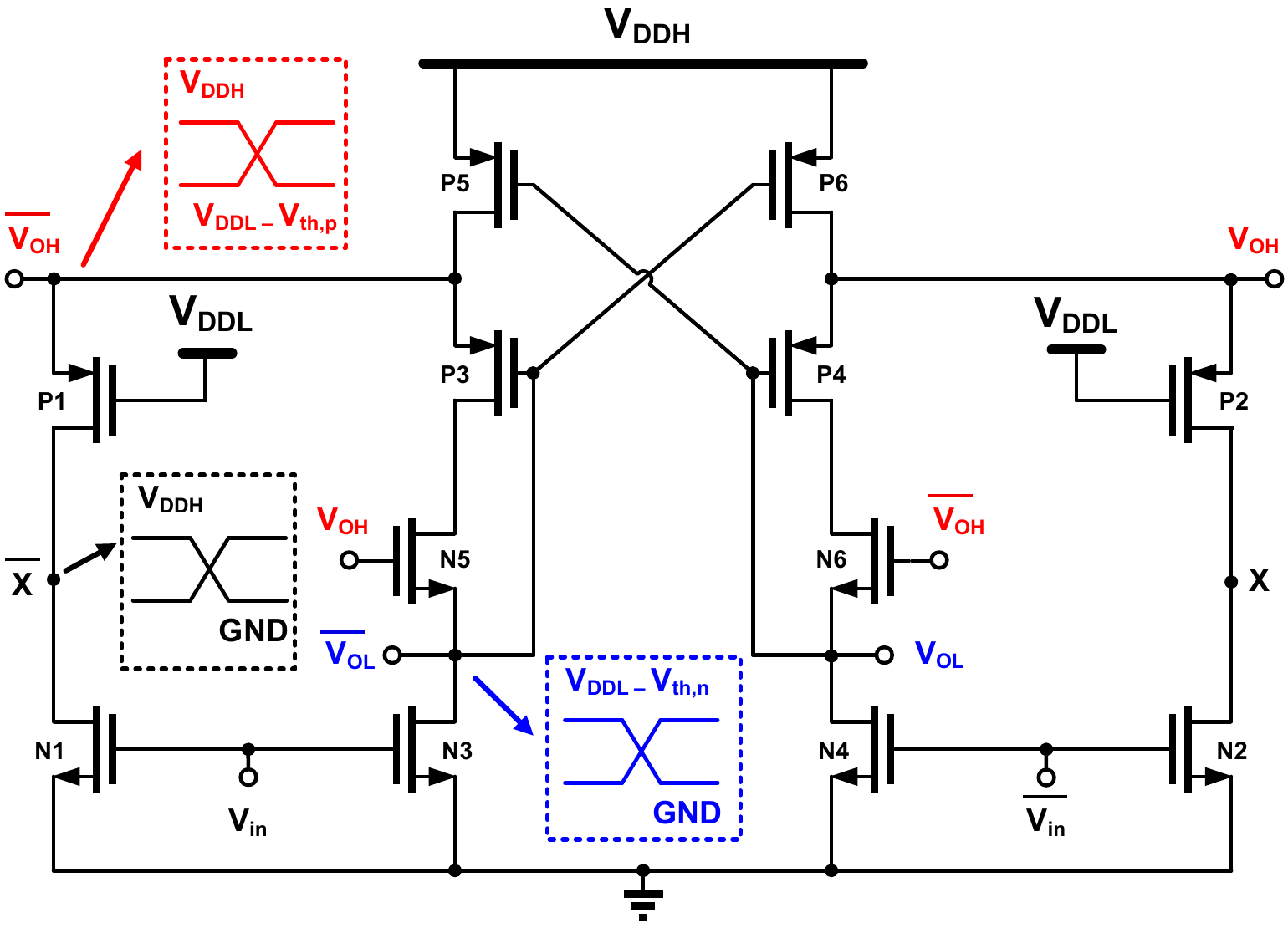}
\caption{Schematic of proposed HVLS.}
\label{fig_ls_prop}

\end{figure}

The schematic and supported voltage transitions of the proposed HVLS are illustrated in \Cref{fig_ls_prop}, and the device sizes are listed in Table~\ref{tab_hvls_size}. To address limitations in earlier designs (Section~\ref{sec_litrev}), the architecture integrates multiple enhancement techniques. A cross-coupled PMOS pair (P5–P6) accelerates output transitions, while diode-connected PMOS devices (P3–P4), placed in series with N5–N6, mitigate current contention. These NMOS devices also serve as cross-coupled transistors, with their gates tied to $\overline{V_{OH}}$ and $V_{OH}$, which improves transition sharpness, by using a regenerative latching effect. Additionally, P1 and P2 limit the contention current in the leftmost and rightmost branches.

\begin{table}[t]
\centering
\caption{Device Sizes (W/L) of the Proposed \ac{HVLS}}
\label{tab_hvls_size}
\resizebox{0.7\columnwidth}{!}{%
\begin{tabular}{|c|c|c|}
\hline
P1-P6    & N1-N4 & N5,N6    \\ \hline
400n/20n & 280n/20n & 140n/20n \\ \hline
\end{tabular}%
}
\end{table}

\begin{figure}[t]
\centering
\includegraphics[width=0.45\textwidth]{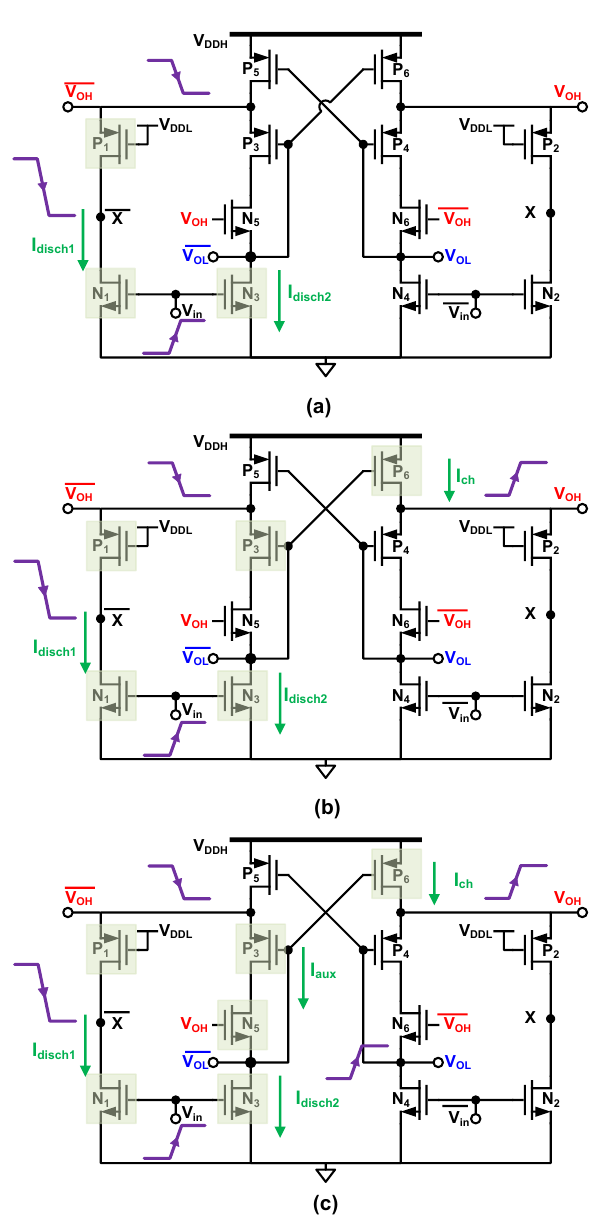}
\caption{\ac{HVLS} operation: (a) activation of current-steering branches $\mathrm{I_{disch1}}$ and $\mathrm{I_{disch2}}$ due to $V_{in}$ transition; (b) P6 and P4 turning on as $\mathrm{\overline{V_{OL}}}$ discharges and $\mathrm{V_{OH}}$ charges; (c) accelerated discharge of $\overline{V_{OH}}$ via $\mathrm{I_{aux}}$.}
\label{fig_ls_prop_op}
\vspace{-5mm}
\end{figure}

The operation of the proposed level shifter can be understood in three phases, illustrated in \Cref{fig_ls_prop_op}. When $V_{in}$ transitions from ground to $V_{DDL}$, transistors $N_1$ and $N_3$ turn on and activate the discharge paths $I_{\mathrm{disch1}}$ and $I_{\mathrm{disch2}}$, pulling down nodes $\overline{X}$ and $\overline{V_{OL}}$, as shown in \Cref{fig_ls_prop_op}(a). Since node $\overline{X}$ is initially at $V_{DDH}$, $N_1$ operates in strong inversion once $V_{in}=V_{DDL}$. During this phase, the discharge of $\overline{V_{OH}}$ is limited to $V_{DDL}-V_{th,p}$ due to the gate bias applied to $P_1$.

As $\overline{V_{OL}}$ is discharged, transistor $P_6$ is activated and provides a charging path from $V_{DDH}$, charging nodes $V_{OH}$ and $X$ toward $V_{DDH}$, as illustrated in \Cref{fig_ls_prop_op}(b). This charging current $I_{\mathrm{ch}}$ initiates the transition of the cross-coupled PMOS pair $P_5$–$P_6$, reinforcing the level shifting process.

As $V_{OH}$ increases, transistor $N_5$ turns on and introduces an auxiliary current $I_{\mathrm{aux}}$ that accelerates the discharge of $\overline{V_{OH}}$, as depicted in Fig.~\ref{fig_ls_prop_op}(c). The regenerative interaction between the cross-coupled devices rapidly forces the internal nodes to their final logic levels.
Consequently, nodes $X$ and $\overline{X}$ swing between ground and $V_{DDH}$, whereas $V_{OH}$ and $\overline{V_{OH}}$ are constrained to the range from $V_{DDH}$ to $V_{DDL}-V_{th,p}$, while $V_{OL}$ and $\overline{V_{OL}}$ range from $V_{DDL}-V_{th,p}$ to ground. A symmetric operation occurs on the complementary path when $\overline{V_{in}}$ transitions high. These reduced internal swings do not reach the class-D devices: in both the test circuit (\Cref{fig_total_chip}(a)) and the intended application, $V_{OL}$ and $V_{OH}$ drive inverter buffers supplied from GND--$V_{DDL}$ and $V_{DDL}$--$V_{DDH}$, respectively, which restore full rail-to-rail swings. Because $V_{DDL}-|V_{th,p}|\approx0.7$~V at $V_{DDL}=0.9$~V is well above the switching threshold of the low-domain buffer ($\approx V_{DDL}/2$), the buffers switch with adequate noise margin; the same holds for the low level of $V_{OH}$ with respect to the high-domain buffer.

To prevent overstress under high supply voltages, the gates of $P_1$ and $P_2$ are biased at $V_{DDL}$. In addition, diode-connected transistors $P_3$ and $P_4$ reduce the effective impedance at nodes $V_{OH}$ and $\overline{V_{OH}}$, improving the switching speed. The regenerative cross-coupled structure introduces a strong negative input resistance during transitions, resulting in fast and low-jitter output switching.

\begin{figure}[t]
\centering
\centerline{\includegraphics[width= 0.5\textwidth]{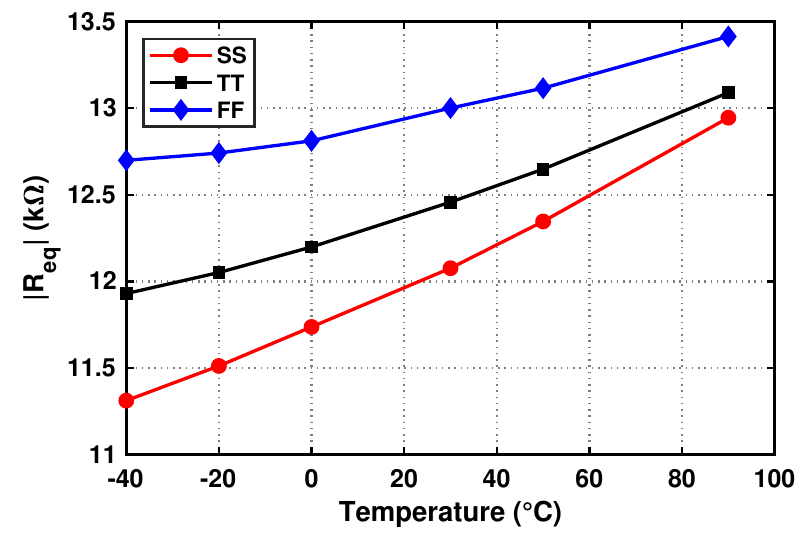}}
\caption{Simulated equivalent negative resistance $|R_{\mathrm{eq}}|$ of the proposed \ac{HVLS} across TT, FF, and SS process corners over the $-40^{\circ}\mathrm{C}$ to $90^{\circ}\mathrm{C}$ temperature range.}
\label{fig_ls_negR}
\vspace{-5mm}
\end{figure}

To analyze this negative resistance and its effect on the performance of the circuit, we define the following substitutions for compactness:
\begin{equation}
    A = g_{m,p3} r_{o,p3},
    \quad B = g_{m,n5} r_{o,n5},
    \quad C = g_{m,p5} r_{o,p5},
\end{equation}
with these definitions, the equivalent resistance can be expressed as:
\begin{equation}
    R_{eq}(A,B) \approx
    -2\times \dfrac{R_0 + r_{o,p5}(A-B)}{A - C - B + CB - AC},
    \label{eq_req_initial}
\end{equation}
where
\begin{equation}
    R_{0} = r_{o,p3}+r_{o,n5}+r_{o,p5},
\end{equation}
the denominator of \eqref{eq_req_initial} can be simplified as
\begin{equation}
    A - C - B + CB - AC
    = (1-C)(A-B) - C.
\end{equation}

Introducing the difference variable
\begin{equation}
    \Delta = A-B,
\end{equation}
\eqref{eq_req_initial} reduces to the fractional-linear form
\begin{equation}
    R_{eq}(\Delta) =
    -2\times \dfrac{R_0 + r_{o,p5}\Delta}{(1-C)\Delta - C},
    \label{eq_req_Delta}
\end{equation}
to have a strong positive feedback in order to reduce the jitter, the denominator should approach zero. Hence, $\Delta$ should be equal to:
\begin{equation}
    \Delta = \dfrac{C}{1-C}.
    \label{eq_delta_c}
\end{equation}

\begin{table*}[b]
\hrule
\begin{equation}
\resizebox{0.93\textwidth}{!}{$\displaystyle
R_{eq} \approx -2\times \frac{r_{o,p_3} + r_{o,n_5}+r_{o,p_5}+g_{m,p_3}\cdot r_{o,p_3}\cdot r_{o,p_5} - g_{m,n_5}\cdot r_{o,n_5}\cdot r_{o,p_5}}{g_{m,p3} \cdot r_{o,p3} - g_{m,p5} \cdot r_{o,p5} - g_{m,n5} \cdot r_{o,n5} + g_{m,p5} \cdot r_{o,p5} \cdot g_{m,n5} \cdot r_{o,n5} - g_{m,p3} \cdot r_{o,p3} \cdot g_{m,p5} \cdot r_{o,p5} }$}
\label{eq_neg_res}
\end{equation}
\end{table*}

The denominator of \eqref{eq_req_Delta} vanishes at $\Delta=C/(1-C)$, where the resistance seen at $N_3$/$N_4$ changes sign; \eqref{eq_delta_c} is thus the condition for the cross-coupled network to regenerate signals. Beyond this threshold $R_{eq}$ is negative and $|R_{eq}|$ decreases as $\Delta$ increases, which shortens the regeneration time constant $\tau_{\mathrm{reg}}\propto C_{\mathrm{node}}|R_{eq}|$ and hence reduces jitter. The devices $N_5$ and $P_3$ are therefore sized to place $\Delta$ well beyond $C/(1-C)$ within the available power and area budget.
The negative resistance seen at N3 and N4 is given by~\eqref{eq_neg_res}, where $\mathrm{g_{m,(p/n)i}}$ and $\mathrm{r_{o,(p/n)i}}$ are the corresponding trans-conductance and channel length modulation resistance of the transistors in the proposed HVLS. Timing jitter at a threshold crossing is, to first order, the voltage noise at the crossing divided by the slope of the waveform there, $\sigma_t \approx \sigma_v/(\mathrm{d}v/\mathrm{d}t)$. In a conventional level shifter the slope is set by the (weak) current of the input devices, so the output dwells near the switching threshold for a comparatively long time, during which noise accumulates. In the proposed HVLS, once the input branches have steered enough current to tip the cross-coupled pair, the node voltages grow exponentially with a regeneration time constant $\tau_{\mathrm{reg}} \propto C_{\mathrm{node}}\,|R_{\mathrm{eq}}|$, where $R_{\mathrm{eq}}<0$ is the resistance derived in~\eqref{eq_neg_res}. The slope at the crossing is therefore sharpened by the regeneration. Because $V_{OL}$ and $V_{OH}$ are both derived from the same regenerative event, their timing errors are strongly correlated, so the \emph{skew} between them is smaller still than the jitter of either output.

\Cref{fig_ls_negR} shows the variation of the equivalent resistance $|R_{\mathrm{eq}}|$ of the proposed HVLS across TT, FF, and SS process corners over a temperature range from $-40^{\circ}\mathrm{C}$ to $90^{\circ}\mathrm{C}$. As observed, $|R_{\mathrm{eq}}|$ increases monotonically with temperature for all process corners. In the SS corner, $|R_{\mathrm{eq}}|$ increases from approximately $11.4~\mathrm{k}\Omega$ to $12.9~\mathrm{k}\Omega$, while in the TT corner it varies from about $12.0~\mathrm{k}\Omega$ to $13.1~\mathrm{k}\Omega$. The FF corner exhibits the highest resistance values, ranging from approximately $12.7~\mathrm{k}\Omega$ to $13.4~\mathrm{k}\Omega$ across the same temperature range.

\begin{figure}[t]
\centering
\centerline{\includegraphics[width= 0.45\textwidth]{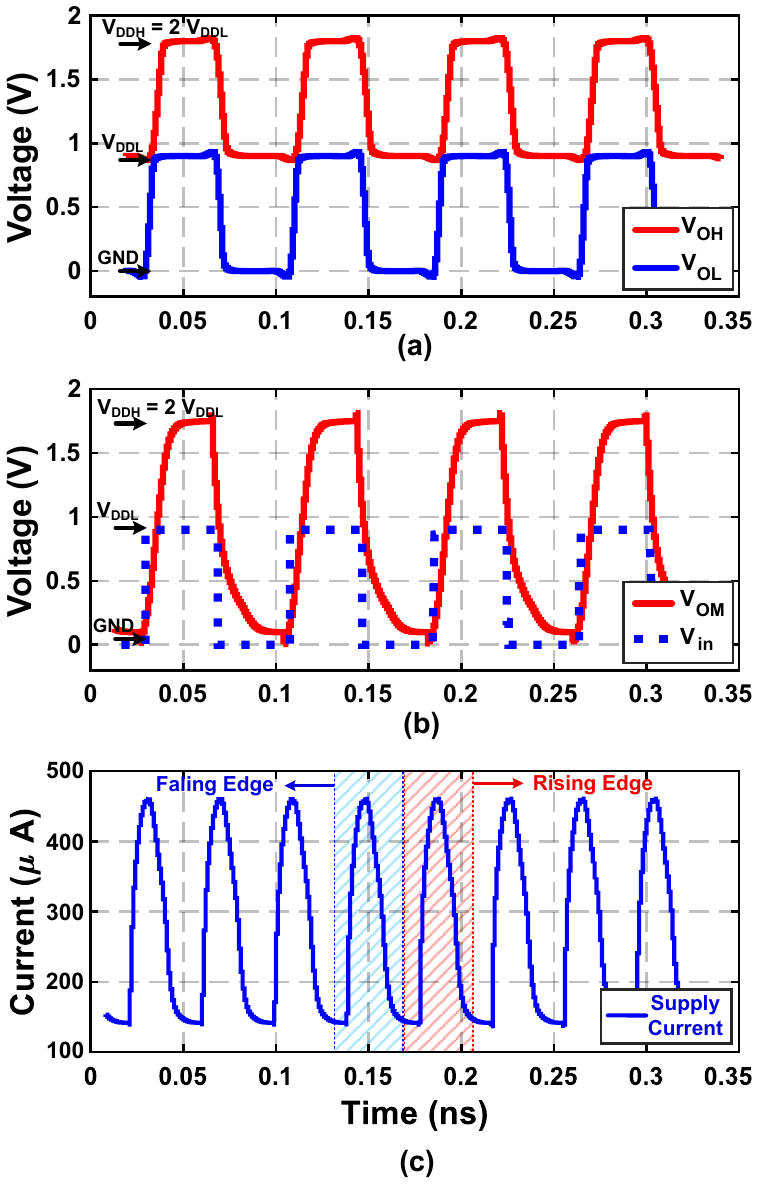}}
\caption{Simulation of (a) $\mathrm{V_{OH}}$ and $\mathrm{V_{OL}}$, (b) output voltages, and (c) supply current waveform.}
\label{fig_ls_trans}
\vspace{-5mm}
\end{figure}

This trend can be attributed to temperature-dependent degradation of carrier mobility and transconductance, which effectively increase the devices' output resistance and, consequently, the circuit's equivalent resistance. Since the effective negative resistance introduced by the positive feedback network is directly related to the regeneration strength, an increase in $|R_{\mathrm{eq}}|$ reduces the regenerative gain. As a result, the switching transition becomes slower and more susceptible to timing uncertainty, which contributes to the output jitter that will be discussed later.

\section{Performance Evaluation of HVLS for Analog and Digital Applications}
\label{sec_sim}

\begin{figure}[t]
\centering
\includegraphics[width=0.45\textwidth]{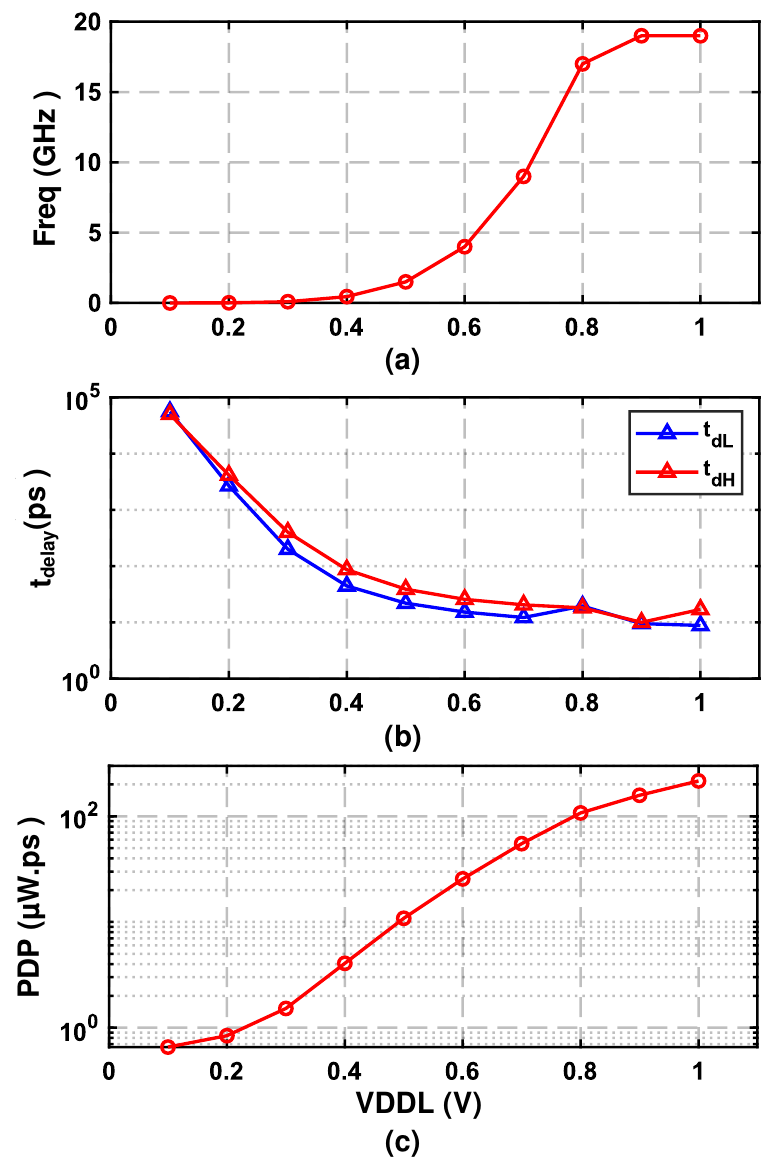}
\caption{(a) Maximum operating frequency, (b) propagation delays, and (c) PDP of HVLS at maximum operation frequency over different supplies ($\mathrm{V_{DDH}}$ = $\mathrm{2V_{DDL}}$).}
\label{fig_ls_wave}
\vspace{-5mm}
\end{figure}

As illustrated in \Cref{fig_ls_SoC}, \acp{VLS} are used in applications ranging from digital I/O transitions to mixed-signal circuits and power management ICs. Accordingly, this section will further evaluate the performance of the proposed \ac{HVLS} for applications, assessing its effectiveness in both purely digital circuits and mixed-domain circuits, such as the aforementioned cascoded class-D amplifiers. Similar to the comparative study of the VLS topologies in Section \ref{sec_litrev}, the proposed HVLS is driving a minimum-sized inverter as a load.

\subsection{Analog Applications Results of the Proposed HVLS}
To corroborate the analytical results, post-layout simulations were performed using a 22-nm FD-SOI process technology. The \ac{HVLS} was excited with the target input waveform, and the corresponding transient responses are illustrated in \Cref{fig_ls_trans}~(a)--(c). The plots show the voltage transitions at the output nodes and the instantaneous current drawn from the supply. The simulation was conducted at 12.2~GHz under a dual-supply condition with $V_{DDH}=2\times~V_{DDL}$, where $V_{DDL}=0.9~\mathrm{V}$ denotes the technology's nominal voltage. At 12.2~GHz the supply current does not fall below approximately 125~$\mu$A, because the internal nodes do not have time to settle fully within each half-period, so neither half of the circuit turns off completely before the next transition. With the input held constant, the static current at $V_{DDL}=0.9$~V and $V_{DDH}=1.8$~V is 83~$\mu$A, i.e., a static power of 150.6~$\mu$W (Table~\ref{VLS_comp_tab_high}); it flows because the $V_{OH}$ nodes settle at $V_{DDL}-|V_{th,p}|$, which does not fully turn off the diode-connected $P_3$/$P_4$.

Throughout this section, $V_{DDH}=2V_{DDL}$, the gates of $P_1$/$P_2$ are tied to $V_{DDL}$, and each output drives a minimum-size inverter, the same load as in Section~\ref{sec_litrev} and in the fabricated circuit. The maximum operating frequency is the highest input frequency at which both outputs still toggle rail-to-rail with a duty cycle of $50\pm1\%$, and propagation delays are measured from the 50\% point of the input transition to the 50\% point of the respective output swing (GND--$V_{DDL}$ for $V_{OL}$, $V_{DDL}$--$V_{DDH}$ for $V_{OH}$). Simulation results in \Cref{fig_ls_wave} confirm high-speed operation, with a supply of $2V_{DDL}$. As shown in \Cref{fig_ls_wave}~(a), the HVLS supports frequencies up to $\approx19~GHz$ at $V_{DDH}$ equal to 1.8~V while the input is toggled between GND and $V_{DDL}$. Lower-supply operation is possible at reduced frequencies, resulting in power savings. \Cref{fig_ls_wave}~(b) displays propagation delays for low-voltage-domain (e.g.,~$V_{OL}$ ($t_{dL}$)) and high-voltage-domain (e.g.,~$V_{OH}$ ($t_{dH}$)) transitions. The \ac{HVLS} exhibits good symmetry and flat delay characteristics above 0.6~V. Moreover, the power-delay product (PDP), plotted in \Cref{fig_ls_wave}~(c), increases with frequency and voltage, as expected.

\begin{figure}[t]
\centering
\centerline{\includegraphics[width= 0.5\textwidth]{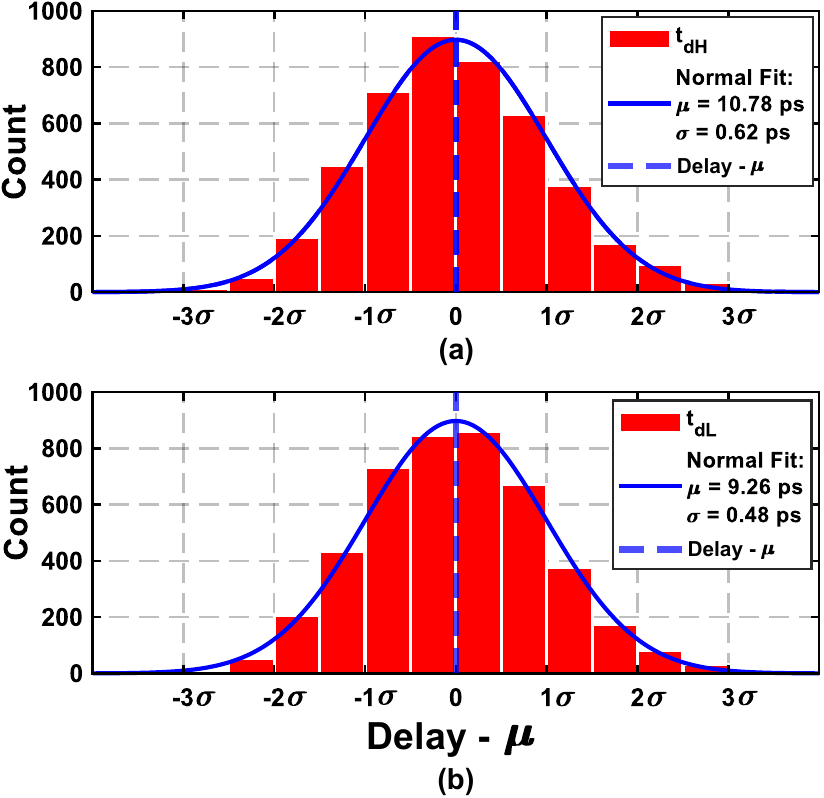}}
\caption{Monte Carlo simulations of the HVLS for (a) $t_{dH}$, (b) $t_{dL}$.}
\label{fig_ls_mont}
\vspace{-5mm}
\end{figure}

For analog performance evaluation and to assess the robustness of the proposed design, 4500 Monte Carlo simulations were performed on the post-layout extracted netlist, with process and mismatch variations enabled simultaneously, at the nominal operating point ($V_{DDL}=0.9$~V, $V_{DDH}=1.8$~V, 12.2~GHz) and with the same minimum-size inverter load. \Cref{fig_ls_mont} illustrates the Monte Carlo simulation results for the proposed circuit's delays from input to both outputs. The mean delay for the $V_{OH}$ ($t_{dH}$) is 10.78 ps with a standard deviation of 0.62 ps. Meanwhile, for the other output $V_{OL}$, the mean delay is 9.26 ps with a standard deviation of 0.48 ps. Considering three times the standard deviation ($3\sigma$), the maximum variation of the delay changes by only 1.86 ps in 99.73\% of the cases. The mean difference between the NFET and PFET driving signals ($\approx1.5$~ps) corresponds to an edge mismatch of $\approx1.85\%$ of the period, which according to Fig.~\ref{fig_LS_pout_tmis} results in less than $0.5\%$ ($<0.02$~dB) output-power degradation of the cascoded class-D PA.

\begin{figure}[t]
\centering
\centerline{\includegraphics[width=0.5\textwidth]{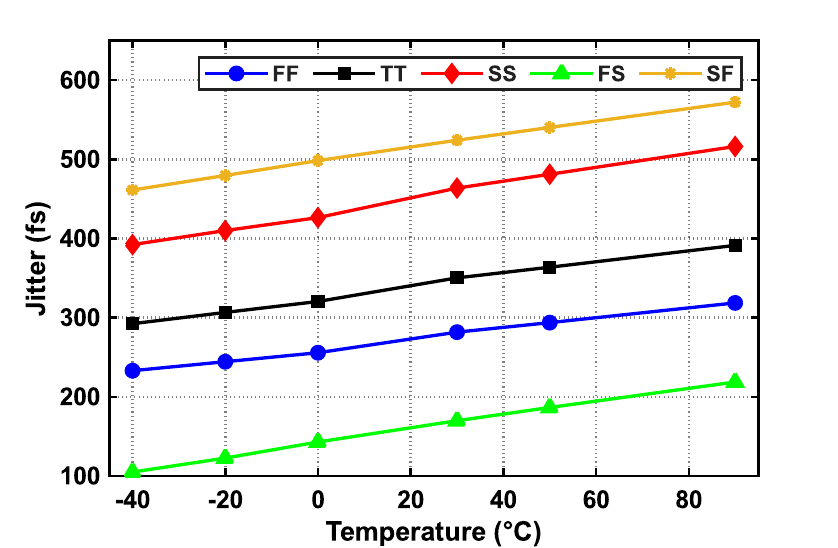}}
\caption{Simulated rms jitter of the proposed \ac{HVLS} at FS, FF, TT, SF, and SS process corners as a function of temperature ($-40^{\circ}\mathrm{C}$ to $90^{\circ}\mathrm{C}$).}
\label{fig_corner_temp}
\vspace{-5mm}
\end{figure}

Process corners and temperature variations have a pronounced impact on circuit performance metrics. Since timing jitter plays a critical role in determining the linearity and spectral purity of digital transmitters and switched-capacitor power amplifiers (SCPAs), it is essential to evaluate the jitter behavior of the proposed HVLS across process and temperature variations. \Cref{fig_corner_temp} illustrates the simulated jitter of the proposed HVLS under TT, FF, and SS corners over a temperature range from $-40^{\circ}\mathrm{C}$ to $90^{\circ}\mathrm{C}$, obtained from the eye diagram of a transient-noise simulation of the stand-alone \ac{HVLS} with a noise bandwidth of 24.4~GHz.

In the FF corner, the jitter increases from approximately $240\,\mathrm{fs}$ at $-40^{\circ}\mathrm{C}$ to about $320\,\mathrm{fs}$ at $90^{\circ}\mathrm{C}$. Similarly, in the TT corner, the jitter rises from around $290\,\mathrm{fs}$ to nearly $400\,\mathrm{fs}$ as temperature increases. A comparable trend is observed in the SS corner, where the jitter varies by approximately $120\,\mathrm{fs}$ across the same range, increasing from $\approx 400\,\mathrm{fs}$ to $\approx 520\,\mathrm{fs}$. The SF corner results in the largest jitter among all evaluated corners, increasing from approximately 460~fs at $-40^\circ C$ to 570~fs at $90^\circ C$. The reduced NMOS transconductance weakens both the current-steering branches and the regenerative feedback, which slows the transitions and increases jitter. In contrast, the FS corner exhibits the lowest jitter, increasing from 100~fs to 220~fs over the same temperature range. In this corner, the pull-down network has higher strength and can discharge the nodes faster due to higher current in the discharge path. Thus, sharper transitions lead to lower output jitter.

This behavior is attributed to the corner-dependent variation of the $R_{\mathrm{eq}}$, as illustrated in \Cref{fig_ls_negR}. In slow-process corners, reduction in device transconductance weakens regenerative feedback, resulting in smaller negative resistance, slower switching transitions, and increased jitter. However, this trend is not valid across temperature variations, since higher temperatures increase noise power, leading to higher output jitter despite improvements in $R_{\mathrm{eq}}$. Conversely, lower temperatures strengthen the negative resistance, which accelerates regeneration, and reduce jitter, as shown in \Cref{fig_corner_temp}. The current-source branches, the most right and the most left branches, in the HVLS introduce additional noise that offsets the benefit of positive feedback, limiting the jitter improvement achievable through $R_{\mathrm{eq}}$ reduction. Consequently, jitter does not scale directly with $R_{\mathrm{eq}}$. Achieving further jitter reduction requires stronger positive feedback; however, such an approach increases power dissipation and area while maintaining the desired operating frequency. Therefore, the proposed architecture represents a trade-off among jitter, speed, power, and area.

\begin{figure}[t]
\centering
\centerline{\includegraphics[width= 0.5\textwidth]{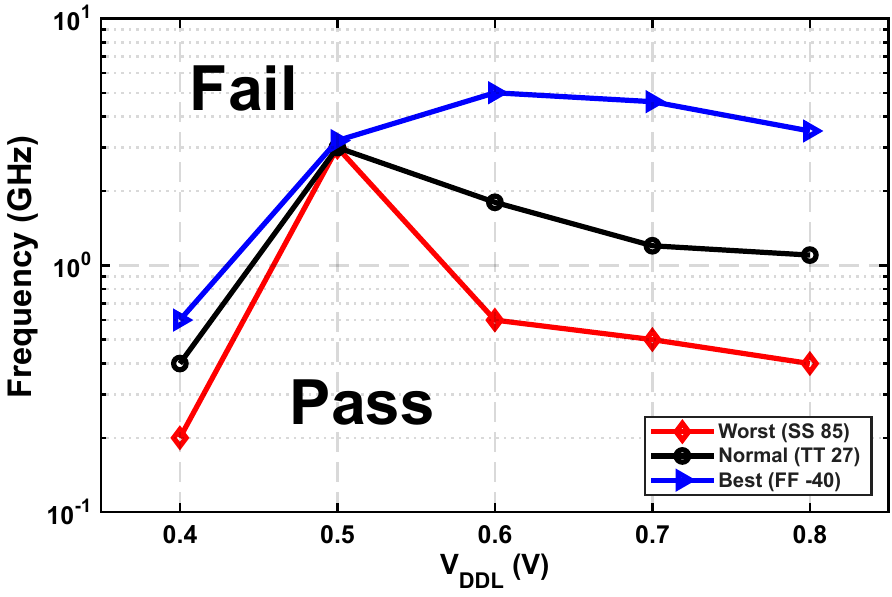}}
\caption{Maximum operation frequency of the HVLS at different input voltages over corners.}
\label{fig_ls_corner_fin}
\vspace{-5mm}
\end{figure}

\begin{figure}[t]
\centering
\centerline{\includegraphics[width= 0.5\textwidth]{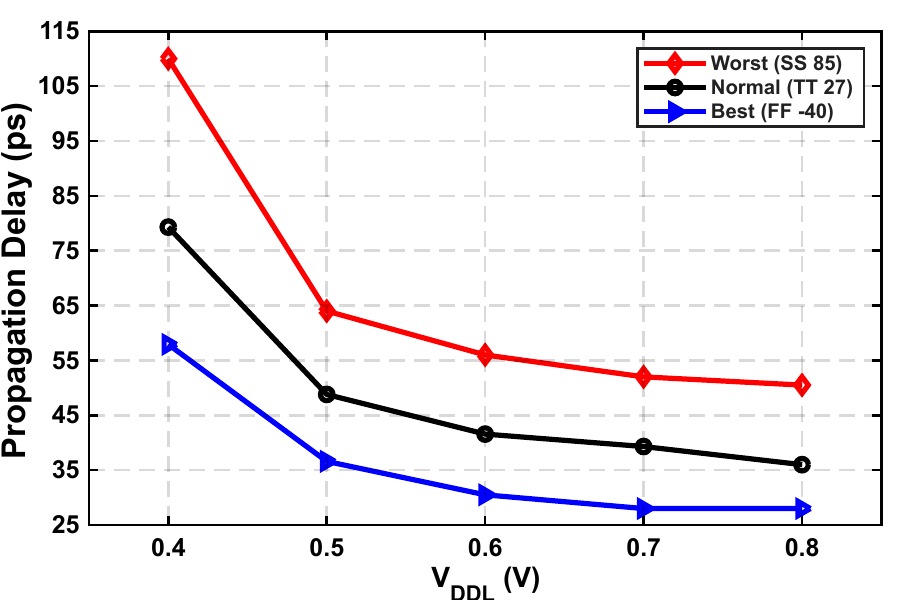}}
\caption{Propagation delay variation of the proposed circuit at different input voltages over corners.}
\label{fig_ls_corner_d}
\vspace{-5mm}
\end{figure}

\begin{figure}[t]
\centering
\centerline{\includegraphics[width= 0.5\textwidth]{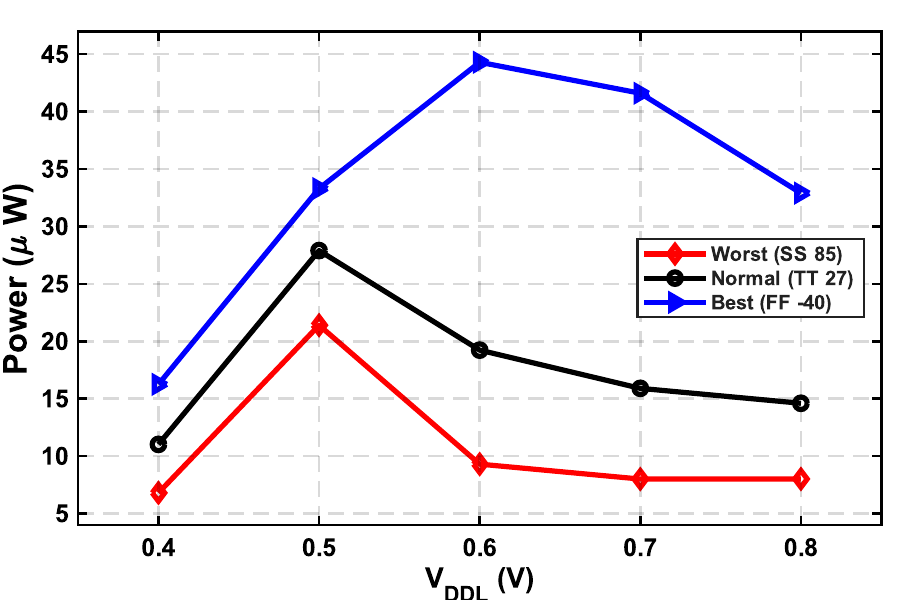}}
\caption{HVLS's power consumption at different input voltages over corners.}
\label{fig_ls_corner_p}
\end{figure}

\begin{figure}[t]
\centering
\centerline{\includegraphics[width= 0.5\textwidth]{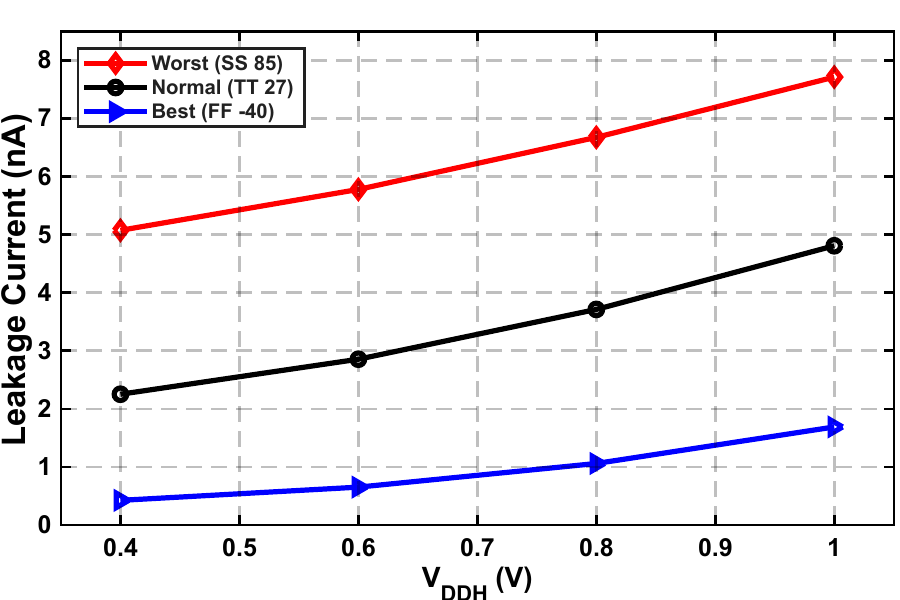}}
\caption{Leakage current of the HVLS at different supply voltages over corners.}
\label{fig_ls_corner_l}
\vspace{-5mm}
\end{figure}

\subsection{Digital Application Results of the Proposed HVLS.}

In digital interfaces, the proposed circuit serves as a high-speed level shifter that translates signal amplitudes from a lower-voltage domain logic ($V_{DDL}$) to a higher-voltage domain ($V_{DDH}$). To evaluate its performance, extensive process corner simulations were conducted. Assuming a constant voltage supply ($V_{DDH}$) equal to the nominal voltage (0.9~V), \Cref{fig_ls_corner_fin} illustrates the maximum operating frequency of the proposed HVLS under different process corners—best (FF, $-40^{\circ}\mathrm{C}$), worst (SS, $85^{\circ}\mathrm{C}$), and Normal (TT, $27^{\circ}\mathrm{C}$)—as input voltage $V_{DDL}$ is varied, mimicking voltage-domain translation. These simulations were performed to isolate the \ac{HVLS}'s supply sensitivity, independent of any preceding circuit stage. The results demonstrate that the proposed design maintains reliable operation up to 3.5~GHz at 0.8~V under the best-case corner (FF, $-40^{\circ}\mathrm{C}$), exceeding the operating frequency of previously reported level shifters discussed in Section~\ref{sec_litrev} by more than a factor of three. The \emph{Pass} region highlighted in \Cref{fig_ls_corner_fin} denotes the frequency range over which the circuit outputs a pulse with $50 \pm 1\%$ duty cycle for the case that input swing is between GND and $V_{DDL}$.

Since the proposed \ac{HVLS} integrates both a current-source stage and a positive-feedback network within its architecture, it exhibits maximum operating frequency when the input swing is in the low-voltage range of 0.4–0.5 V. In this region, transistors N1-N4 remain in strong saturation throughout the entire period when the input signal is high. As the input voltage range increases, however, the duration for which N1-N4 stay in saturation decreases, and these devices enter the triode region after the output transition is completed. In addition, increasing $V_{DDL}$ weakens the discharge paths on both sides of the \ac{HVLS} by limiting their current. Consequently, the circuit's overall maximum operating speed is reduced.

Moreover, \Cref{fig_ls_corner_d} and~\ref{fig_ls_corner_p} respectively illustrate the propagation delay and power consumption of the designed HVLS in the main three corners and conditions relative to the input value. Compared to \Cref{fig_comp_pd}, the design has less delay, especially at high frequencies. Less delay is achieved due to the two positive feedbacks employed in this circuit, which accelerate the transition and improve signal sharpness. Finally, the leakage current of the circuit, as a function of the supply voltage, is shown in \Cref{fig_ls_corner_l}. At a supply voltage of 1~V, the leakage current of the proposed circuit is approximately 5~nA under normal conditions, which is similar to that of the \ac{DCVS}.

\section{Measurement Results}
\label{sec_res}

\begin{figure*}[t]
\centering
\centerline{\includegraphics[width=\textwidth]{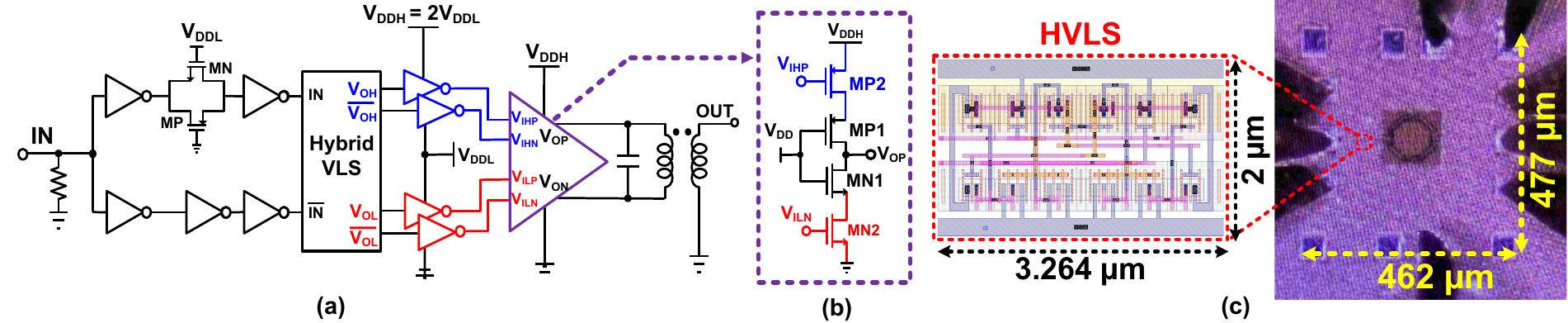}}
\caption{(a) Schematic of HVLS with input and output drivers, (b) stacked class-D pad driver, and (c) chip micrograph.}
\label{fig_total_chip}
\end{figure*}

\begin{figure}[t]
\centering
\centerline{\includegraphics[width=0.45\textwidth]{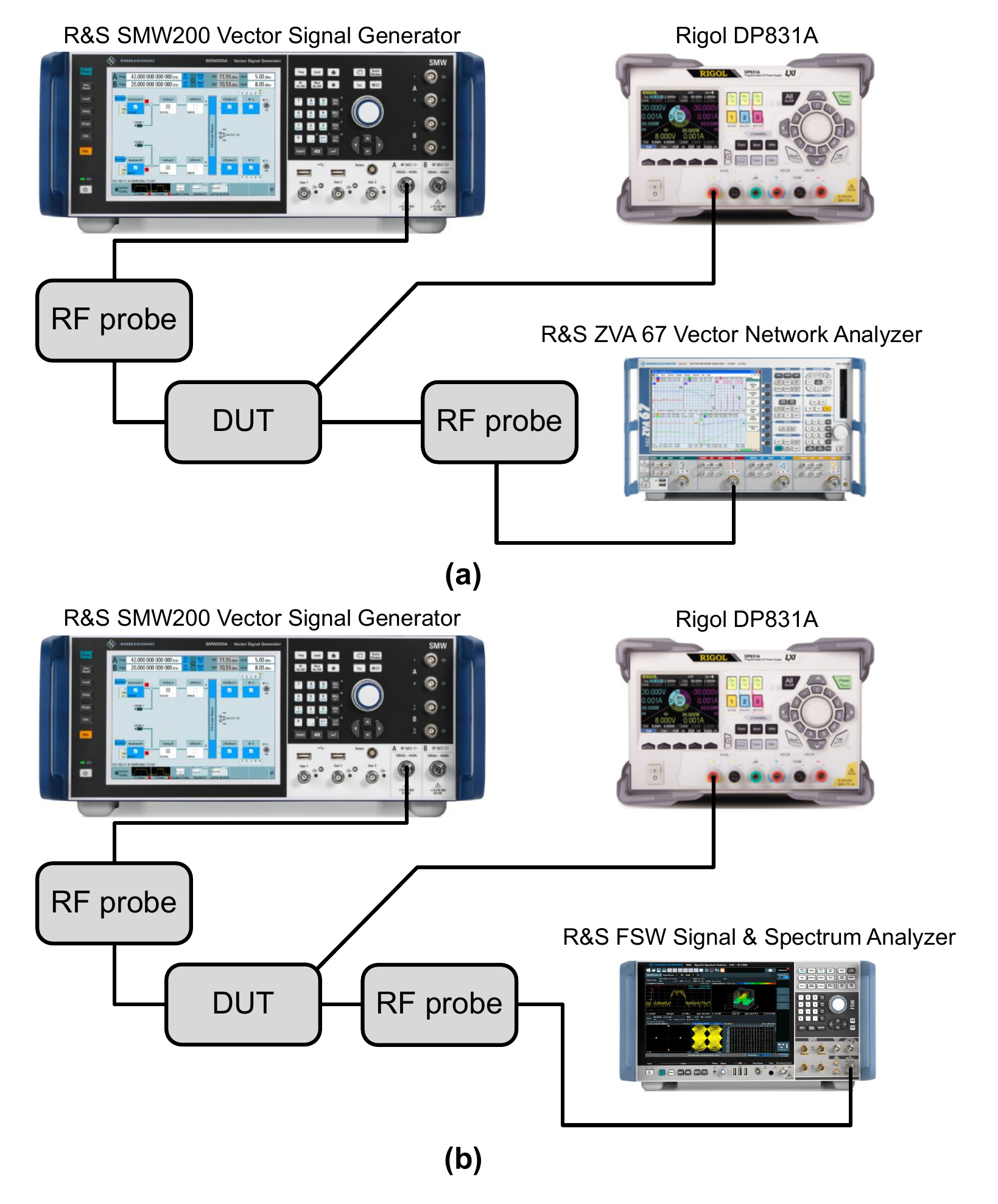}}
\caption{Measurement set up for (a) S-parameter, (b) output spectrum, and phase noise.}
\label{fig_measurement_setup}
\vspace{-5mm}
\end{figure}

The \ac{HVLS} prototype was fabricated in 22-nm FD-SOI process technology. The circuit cannot be measured directly at high frequencies due to the difficulty of driving a low impedance (e.g., $50~\Omega$). Hence, to enable off-chip testing, a single-slice cascoded differential class-D pad driver amplifier with a broadband impedance-matching transformer circuit was designed to demonstrate the \ac{HVLS}'s capabilities. The transformer enabled broadband impedance matching with galvanic isolation, providing ESD protection and a high-frequency direct interface to the I/O pad.

The schematic of the complete test circuit is shown in \Cref{fig_total_chip}~(a). The input RF signal is split into two inverter paths to generate differential pulses, with a transmission gate in the positive path for synchronization. The inverter-based structure ensures fast transitions between ground and $\mathrm{V_{DDL}}$. The \ac{HVLS} outputs drive the aforementioned differential class-D driver. The half-circuit of the cascoded class-D driver is shown in \Cref{fig_total_chip}~(b).

It is important to note that this amplifier is not designed to deliver high output power; it is designed to mimic the typical load that a single \ac{HVLS} would see in an application such as \ac{SCPA}. The cell layout and the chip photograph of the prototype are shown in \Cref{fig_total_chip}~(c). The total chip area, including pads, is 477$\times$462~$\mu$m$^{2}$ while the \ac{HVLS} itself occupies only 3.264$\times$2~$\mu$m$^{2}$.

\begin{figure}[t]
\centering
\centerline{\includegraphics[width= 0.45\textwidth]{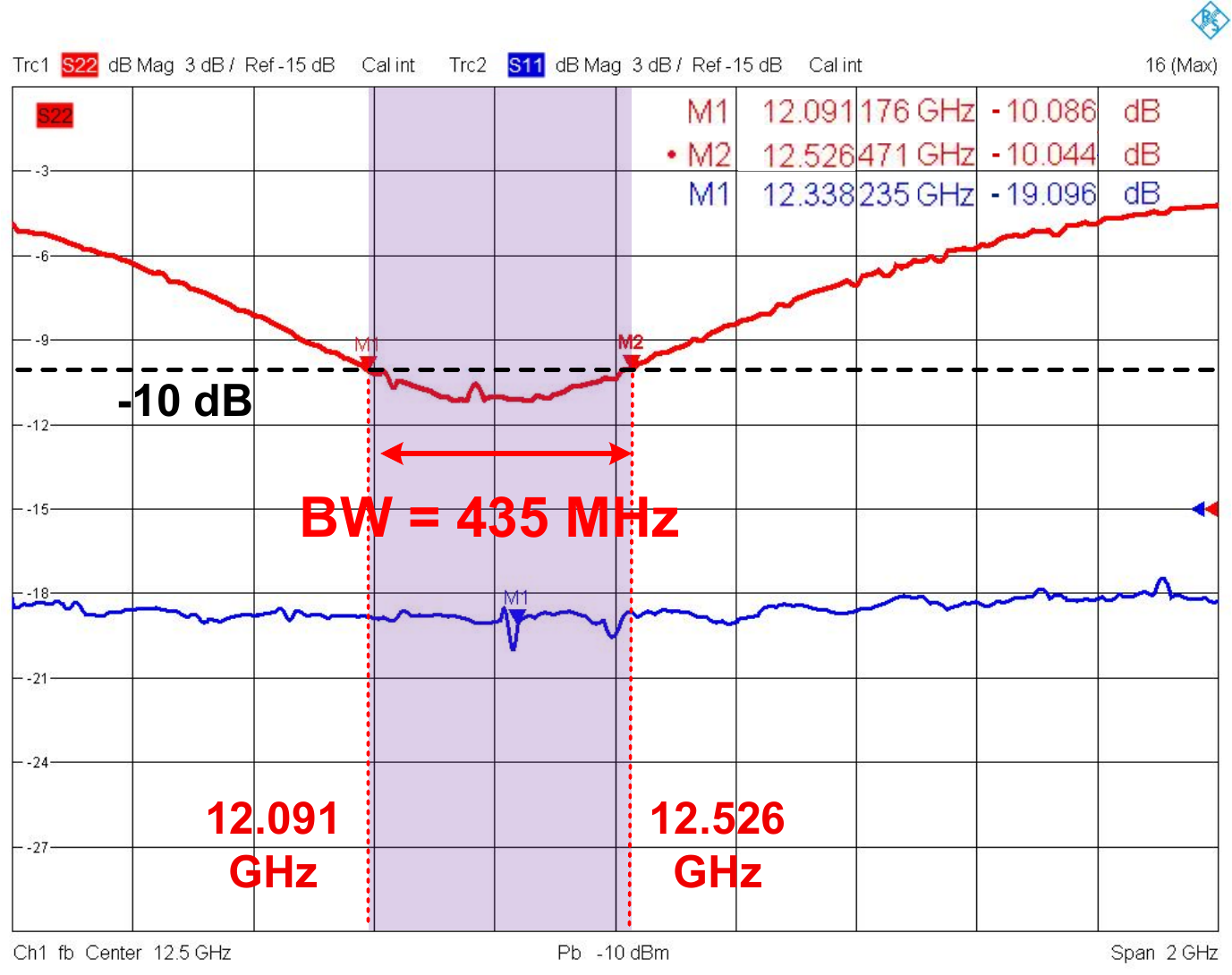}}
\caption{Input and output matching performance.}
\label{fig_mathinc_matching}
\end{figure}

\begin{figure}[t]
\centering
\centerline{\includegraphics[width= 0.45\textwidth]{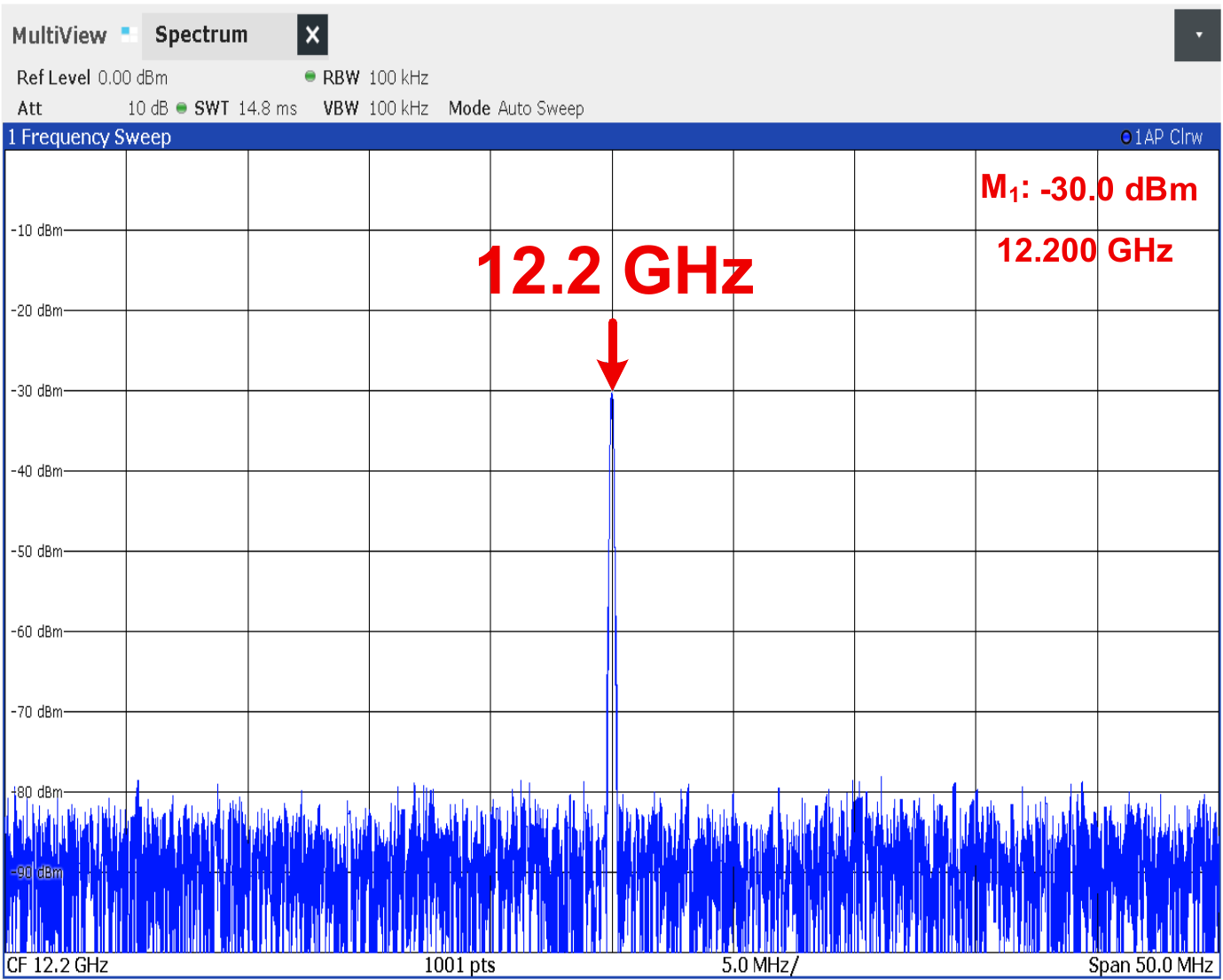}}
\caption{Measured output spectrum.}
\label{fig_mathinc_spectrum}
\vspace{-5mm}
\end{figure}

All measurements were performed at room temperature (27~$^{\circ}$C) on three probed dies, at the nominal supplies $V_{DDL}=0.9$~V and $V_{DDH}=1.8$~V and at $V_{DDH}=1.62$ and 1.98~V ($\pm10\%$). Although the transformer at the output limits the measurement bandwidth to the intended operating frequency, simulations confirm that the circuit can function beyond this range (up to 19~GHz). In order to measure the matching of the circuit, a Rohde \& Schwarz ZVA 67 Vector Network Analyzer is used, as shown in Fig. \ref{fig_measurement_setup}~(a). The input-output matching performance is shown in \Cref{fig_mathinc_matching}. $\rm S_{11}$ is wideband due to the resistive matching, and $\rm S_{22}$ is less than $-$10~dB from 12.1 to 12.5~GHz. Additionally, \Cref{fig_mathinc_spectrum} shows the measured output spectrum of the circuit operating at 12.2~GHz, where the devices used for measurement are shown in \ref{fig_measurement_setup}~(b). As previously noted, the low output power is due to a single-slice output stage, and it matches the simulated predictions. To achieve higher output power in a class-D \ac{PA} or \ac{SCPA}, many of the individual slices used in this design could be combined in parallel, as has been demonstrated previously \cite{ref_scpa_2011}.

\begin{figure}[t]
\centering
\centerline{\includegraphics[width= 0.48\textwidth]{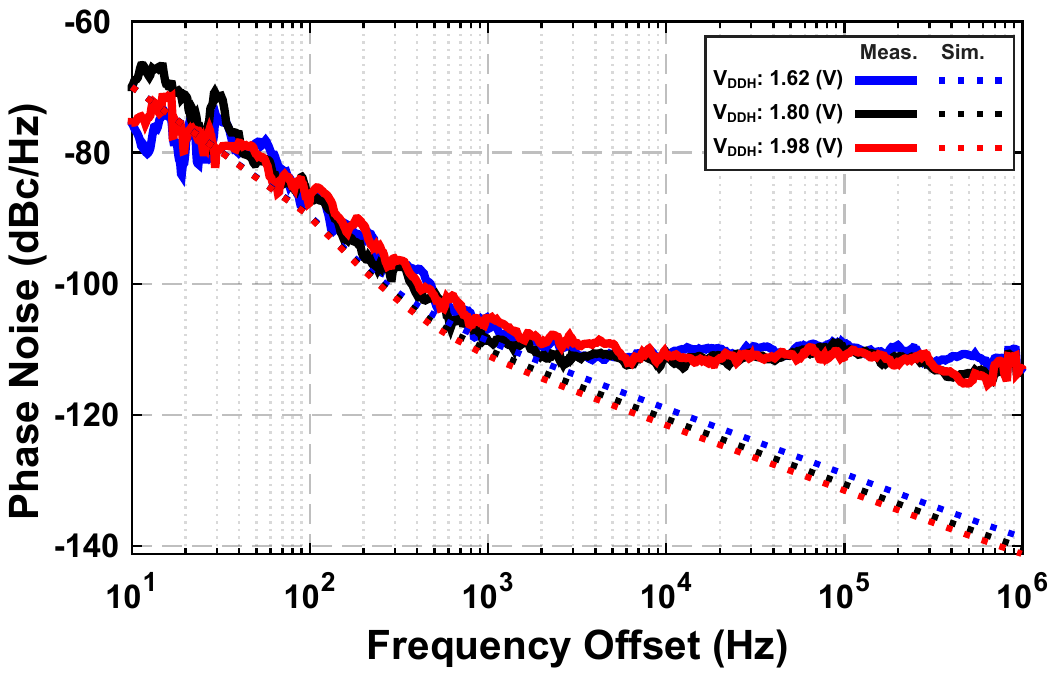}}
\caption{Measured and simulated phase noise at 12.2~GHz for $V_{DDH}=1.62$, 1.8, and 1.98~V. Above a few kilohertz offset, the measurement is limited by the noise floor of the analyzer.}
\label{fig_pn}
\vspace{-5mm}
\end{figure}

\begin{figure}[t]
\centering
\centerline{\includegraphics[width= 0.45\textwidth]{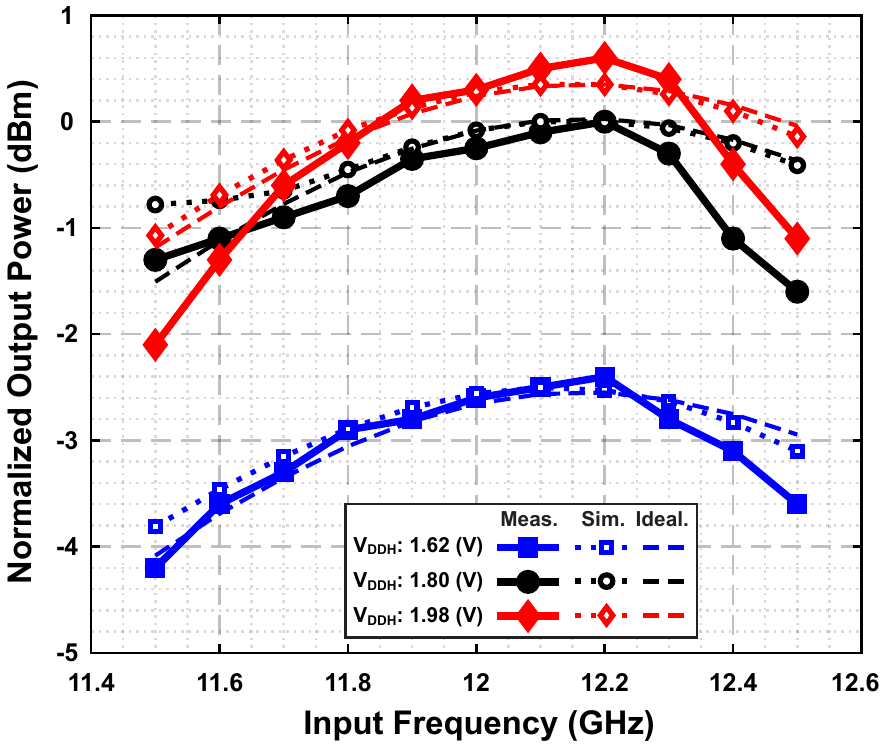}}
\caption{Normalized measured output power over different frequencies with 10\% supply variation.}
\label{fig_output}
\end{figure}

The class-D driver in the output operates if the level shifter can provide proper signals with the required voltage levels. The phase noise (PN) was measured with the phase-noise measurement function of the Rohde \& Schwarz FSW67 signal and spectrum analyzer, as depicted in Fig. \ref{fig_measurement_setup}~(b). Fig. \ref{fig_pn} shows the measured PN of the proposed \ac{HVLS} followed by the cascoded class-D PA both based on simulation and measurement under $\pm 10\%$ supply variation. The rms jitter was obtained by integrating the single-sideband phase noise from 10~Hz to 500~kHz offset. At 12.2~GHz and $V_{DDH}=1.8$~V, the result is 43~fs; under $\pm10\%$ supply variation, it is 38~fs at 1.98~V and 44.5~fs at 1.62~V. Because the output power of the single-slice prototype is low (about $-30$~dBm), the measured phase noise reaches the noise floor of the analyzer, approximately $-110$~dBc/Hz, at offsets above a few kilohertz, whereas the simulated phase noise of the circuit continues to decrease (\Cref{fig_pn}). A measurement of the signal generator alone, with the same analyzer settings and input level, shows the same floor. The measured values are therefore upper bounds set by the measurement setup, and the jitter of the \ac{HVLS} itself is lower; extending the integration would add only instrument noise. Within this limit, the measured jitter varies by less than 10~fs over the $\pm10\%$ supply range. Moreover, \Cref{fig_output} illustrates the implemented circuit's normalized output power as a function of a 10\% adjustment to the supply ($V_{DDH}$). As expected, the output power decreases by 3~dB within the output matching bandwidth. The figure shows the normalized output power of the cascoded class-D PA driven by ideal signals with zero mismatch, the post-layout simulation of the prototype (dotted lines), and the measurement (solid lines). The difference between the ideal and post-layout output power is small; the measurement deviates from the simulation near the edges of the frequency range because of the limited matching of the prototype. It is important to note that the proposed \ac{HVLS} can operate at higher frequencies with wider bandwidths; however, since this case is a stand-alone circuit, not a full \ac{SCPA}, the output bandwidth was limited by the matching network to test and validate its operation.

\begin{figure}[t]
\centering
\centerline{\includegraphics[width= 0.45\textwidth]{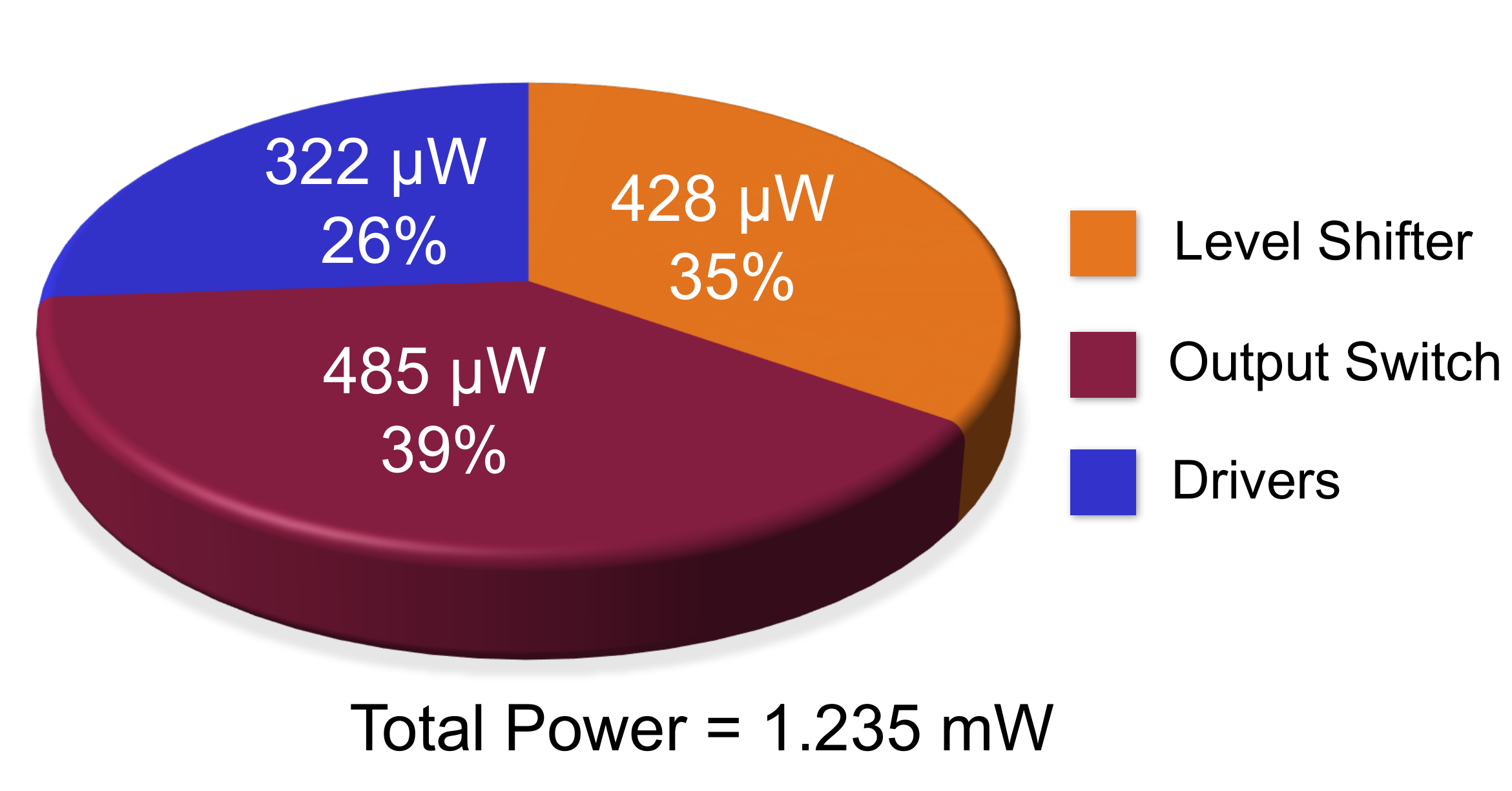}}
\caption{Power breakdown of the total circuit.}
\label{fig_powerBreak}
\vspace{-5mm}
\end{figure}
The power consumption breakdown of the prototype design, based on simulation and measurement, is illustrated in \Cref{fig_powerBreak}. As shown, only 35$\%$ of the power consumption is due to the HVLS, which consumes 427.8~$\mu$W. The balance of power is consumed in the input driver (26\%) and the class-D output pad driver (39\%).

Table~\ref{VLS_comp_tab_low} summarizes the performance of the proposed voltage level shifter and compares it with recently reported state-of-the-art designs for low-power and low-frequency applications. The compared works cover a wide range of CMOS technologies from 130~nm to 7~nm and target different voltage-conversion ranges, operating frequencies, and application requirements. At the low-voltage operating point, the circuit achieves a delay of 3.51~ns with an energy consumption of 12.77~fJ/transition at 1~MHz. Compared with previously reported low-voltage level shifters, the proposed HVLS achieves comparable energy per transition and silicon area, and a shorter propagation delay than six of the nine designs in Table~\ref{VLS_comp_tab_low}. Its static power consumption of 8.26~nW at $V_{DDL}=0.25$~V is higher than that of most of these designs, which are optimized primarily for leakage. The core layout occupies less than $7~\mu\text{m}^2$, excluding input drivers, pads, and the impedance matching network.

For high-speed operation, Table~\ref{VLS_comp_tab_high} compares the proposed HVLS with state-of-the-art level shifters characterized at 400~MHz and above. Relative to prior low-voltage level shifters, the proposed design maintains competitive energy efficiency and area while extending the operating frequency into the multi-GHz regime. The HVLS achieves a propagation delay of 12~ps at 12.2~GHz with an energy consumption of 35.07~fJ per transition. As summarized in Table~\ref{VLS_comp_tab_high}, previously reported high-frequency level shifters have not demonstrated RF-speed voltage-domain conversion in the GHz range. Therefore, the proposed design provides a compact and energy-efficient solution for high-speed level shifting and voltage-domain interfacing.

Overall, the comparison results indicate that the proposed HVLS is suitable for both ultra-low-power and high-frequency applications. While maintaining competitive power, energy, and area metrics, its primary advantage lies in high-speed operation, where it enables GHz-range voltage-domain conversion with substantially lower propagation delay than existing state-of-the-art solutions. Consequently, the proposed architecture is particularly well suited for next-generation high-speed and RF-integrated systems requiring efficient cross-domain signaling.

\section{Conclusion}
\label{sec_conc}

\begin{table*}[t]
\centering
\caption{Performance Summary and Comparison With State-of-the-Art Voltage Level Shifters for Low Frequency}
\label{VLS_comp_tab_low}
\renewcommand{\arraystretch}{1.12}
\setlength{\tabcolsep}{3.0pt}
\scriptsize
\resizebox{\textwidth}{!}{%
\begin{tabular}{c c c c c c c c c c c}
\hline
\textbf{Ref.} &
\cite{ref2} & \cite{ref_LSe} & \cite{ref9} &  \cite{ref10} &
 \cite{ref_comp3} &  \cite{ref_comp4} &  \cite{ref_ls1_2025} &
 \cite{ref_ls_2026} & \cite{ref_ls_tcas2} & \textbf{This Work} \\
\hline

\textbf{Year}
& 2024 & 2023 & 2021 & 2018 & 2023 & 2019 & 2026 & 2026 & 2026 & \textbf{} \\

\textbf{Journal}
& TCAS-II & JSSC & TCAS-II & JSSC-L & TVLSI & TCAS-II & TCAS-I & JSSC & TCAS-II & \textbf{} \\

\textbf{Technology}
& 55 nm & 55 nm & 65 nm & 40 nm & 7 nm & 28 nm FD-SOI & 28~/~7 nm & 28 nm & 130 nm & \textbf{22 nm FD-SOI} \\

\textbf{$V_{DDL}$ (V)}
& 0.3 & 0.3 & 0.2 & 0.3/0.4 & 0.4 & 0.2 & 0.3~/~0.28 & 0.28 & 0.2 & \textbf{0.25} \\

\textbf{$V_{DDH}$ (V)}
& 1.2 & 1.2 & 1.2 & 1.1 & 1.2 & 1.0 & 1.2~/~1.0 & 1.0 & 1.2 & \textbf{0.5} \\

\textbf{Delay (ns)}
& 12.1 & 20.08 & 26.75 & 80 & 0.21 & 10.1 & 10.9~/~0.28 & 0.474 & 1.19 & \textbf{3.51} \\

\textbf{$V_{DDL,min}$ (V)}
& 0.13 & 0.196 & 0.12 & 0.05 & 0.14 & 0.043 & 0.227~/~0.17 & 0.28 & 0.2 & \textbf{0.25} \\

\textbf{@Freq}
& 100 kHz & 1 MHz & 1 MHz & 1 MHz & 1 MHz & N/A & 1 MHz & 10 MHz & 40 MHz & \textbf{1 MHz} \\

\textbf{Energy/Trans. (fJ)}
& 16.13 & 18.11 & 147 & 4.2 & 20.43 & 5.2 & 8.05~/~7.64 & 0.0743 & 16.2 & \textbf{12.77} \\

\textbf{@Freq}
& 1 MHz & 1 MHz & 10 MHz & 1 MHz & 1 MHz & 1 MHz & 1 MHz & 10 MHz & 40 MHz & \textbf{1 MHz} \\

\textbf{Area ($\mu$m$^2$)}
& 7.94 & 9.98 & 6.9 & 8 & 0.418 & 16.6 & 2.37~/~0.204 & 2.37 & 49.4 & \textbf{6.528$^{*}$} \\

\textbf{Static Power}
& 16.72 pW & 0.12 nW & 2.66 nW & 0.55 nW & 7.7 nW & 6.5 nW & 0.33~/~5.04 nW & 13.8 nW & 7.2 nW & \textbf{8.26 nW} \\

\hline
\end{tabular}%
}
\begin{flushleft}
\footnotesize
$^{*}$ HVLS chip area without drivers, pads, and matching network.
\end{flushleft}
\label{table_comp}
\end{table*}

\begin{table*}[t]
\centering
\caption{Performance Summary and Comparison With State-of-the-Art Voltage Level Shifters for High Frequency}
\label{VLS_comp_tab_high}
\renewcommand{\arraystretch}{1}
\setlength{\tabcolsep}{3.0pt}
\scriptsize
\resizebox{0.7\textwidth}{!}{%
\begin{tabular}{c c c c c c c}
\hline
\textbf{Ref.} &
\cite{ref9} & \cite{ref_ls_garcia} & \cite{ref_ls_garcia_2} &  \cite{ref_ls_Nestor}  & \cite{ref_ls_tcas2} & \textbf{This Work} \\
\hline

\textbf{Year}
& 2021 & 2009 & 2019 & 2020 & 2026 & \textbf{} \\

\textbf{Publication Title}
& TCAS-II & ISCIT & MWSCAS & ISCAS & TCAS-II & \textbf{} \\

\textbf{Technology}
& 65 nm & 65nm & 65nm & 180nm & 130nm & \textbf{22 nm FD-SOI} \\

\textbf{$V_{DDL}$ (V)}
& 0.37 & 0.5 & 0.4 & 1.7 & 0.4 & \textbf{0.9} \\

\textbf{$V_{DDH}$ (V)}
& 1.2 & 1 & 0.8 & 3.3 & 1.2 & \textbf{1.8} \\

\textbf{Delay (ns)}
& 1.46 & 0.395 & 0.98 & 1.4 & 1.19 & \textbf{0.012} \\

\textbf{$V_{DDL,min}$ (V)}
& 0.37 & 0.5 & 0.4 & 1.7 & 0.34 & \textbf{0.9} \\

\textbf{@Freq}
& 500 MHz & 500 MHz & 500 MHz & 400MHz & 500 MHz & \textbf{12.2 GHz} \\

\textbf{Energy/Trans. (fJ)}
& N/A & 2540 & 470 & 470 & 16.2 & \textbf{35.07} \\

\textbf{@Freq}
& 500 MHz & 500 MHz & 500 MHz & 1 MHz & 500 MHz & \textbf{12.2 GHz} \\

\textbf{Area ($\mu$m$^2$)}
& 6.9 & 56.55 & 83.76 & 162 & 49.4 & \textbf{6.528$^{*}$} \\

\textbf{Static Power}
& 12.286~$\mu$W & N/A & 240~$\mu$W & 0.99~nW & 7.2~nW & \textbf{150.6~$\mu$W} \\

\hline
\end{tabular}%
}
\begin{flushleft}
\footnotesize
$^{*}$ HVLS chip area without drivers, pads, and matching network.
\end{flushleft}
\label{table_comp2}
\vspace{-5mm}
\end{table*}

This paper presents a novel \ac{HVLS} that simultaneously solves two key problems in high-speed mixed-voltage systems. First, it generates a level-shifted output ($V_{\mathrm{DDH}} = 2V_{\mathrm{DDL}}$) that is inherently synchronous with a second output at the input voltage level ($V_{\mathrm{DDL}}$), eliminating the timing mismatch that arises when separate circuits produce the two domain signals. This synchronicity directly removes the need for programmable delay-calibration chains in cascoded class-D PA applications, which are otherwise required to compensate for PVT-dependent skew between the PMOS and NMOS drive signals. Second, the regenerative cross-coupled feedback network achieves a step-change in operating speed, enabling simulated operation up to 19~GHz—substantially beyond the multi-GHz barrier of prior-art level shifters—thereby making the \ac{HVLS} suitable for \ac{SCPA} and digital transmitters in emerging FR3-band systems. To contextualize the timing-mismatch problem, an analytical model is also developed to quantify the resulting output-power degradation in cascoded class-D PAs. The prototype, fabricated in a 22-nm FD-SOI process technology, is characterized at 12.2~GHz, dissipating 427.8~$\mu$W with a measured rms jitter below 43~fs, limited by the measurement setup, and an energy per transition of 35.07~fJ at 12.2~GHz.

\section{Acknowledgment}
The authors would like to acknowledge the National Science Foundation (NSF) under grant \#2314813 and GlobalFoundries' university partner program for their support.

\section{Appendix A}
\appendices
\label{app_fourier_derivation}

This Appendix provides a detailed derivation of the expressions used in Section~III for calculating the output power degradation in a cascoded class-D power amplifier (PA) due to delay mismatch between the PMOS and NMOS switching signals. The analysis is divided into two parts. First, the ideal switching case is analyzed to derive the output power using the fundamental Fourier component of the output waveform. Second, the non-ideal waveform transitions caused by device parasitics are modeled using ramp and exponential segments whose Fourier coefficients are derived.

\subsection{Ideal Cascoded Class-D PA}

In the ideal case, the switches are assumed to have negligible on-resistance and zero parasitic capacitance. Consequently, the output waveform transitions instantaneously between the supply rails. When the PMOS and NMOS switching signals are perfectly synchronized, the output voltage waveform is a two-level square waveform with amplitude $2V_{DD}$ and a duty cycle of $50\%$, as illustrated in Fig.~\ref{fig_ls_mm_wave_ideal}~(a).

\subsubsection{Output Power of the Ideal Class-D PA}

Since the RF output power is determined by the fundamental component delivered to the load resistance $R_{opt}$, the output power is \cite{ref_scpa_2011}:

\begin{equation}
P_{out,i}=\frac{8}{\pi^2}\frac{V_{DD}^2}{R_{opt}},
\label{eq_pout}
\end{equation}
This expression represents the ideal output power of the cascoded class-D PA.

\subsection{Effect of Delay Mismatch}

In practice, the gate signals of the PMOS and NMOS devices may exhibit a delay mismatch, denoted by $T_{ms}$, which modifies the effective pulse width of the output waveform. Defining the half-period of the switching waveform as

\begin{equation}
T_c=\frac{1}{2f_c},
\end{equation}

the normalized delay mismatch is given by

\begin{equation}
\delta=\frac{T_{ms}}{2T_c}=\frac{T_{ms}}{T}.
\end{equation}

Due to this mismatch, the duration of the full output level, $2V_{DD}$, is reduced from $T_c$ to $T_c-T_{ms}$, while an additional intermediate voltage step is introduced during the mismatch interval. Consequently, the fundamental Fourier component of the output waveform is modified. Evaluating the Fourier series, considering the time mismatch, the resulting nonideal waveform yields

\begin{equation}
P_{out,ni} =
P_{out,i}
\left[
\sin^2\left(\pi\left(\frac{1}{2}-\delta\right)\right)
+
\frac{1}{4}
\sin^2\left(\pi\delta\right)
\right].
\end{equation}

The first term accounts for the duration of the fundamental component caused by the shortened $2V_{DD}$ pulse, whereas the second term represents the contribution of the mismatch-induced intermediate voltage step. The factor $1/4$ in the second term follows directly from the corresponding Fourier coefficient. For a rectangular pulse of amplitude $A$ and duration $\tau$ within a period $T$, the magnitude of the fundamental Fourier coefficient is

\begin{equation}
|C_1|
=
\frac{A}{\pi}
\left|
\sin\left(\pi\frac{\tau}{T}\right)
\right|.
\end{equation}

Since the mismatch-induced step has an amplitude of $V_{DD}$, whereas the main output pulse has an amplitude of $2V_{DD}$, its normalized fundamental component is

\begin{equation}
\frac{|C_{1,ms}|}{|C_{1,i}|}
=
\frac{V_{DD}}{2V_{DD}}
\sin\left(\pi\delta\right)
=
\frac{1}{2}
\sin\left(\pi\delta\right).
\end{equation}

Since the output power is proportional to the squared magnitude of the Fourier coefficient, the mismatch-induced contribution becomes:

\begin{equation}
\left|
\frac{C_{1,ms}}{C_{1,i}}
\right|^2
=
\frac{1}{4}
\sin^2\left(\pi\delta\right).
\end{equation}
The two contributions can be added in power because the shortened $2V_{DD}$ pulse, of width $T_c-T_{ms}$, and the mismatch-induced step, of width $T_{ms}$ immediately following it, are centred a quarter period apart; their fundamental components are therefore in quadrature, and the cross-term in $|C_1|^2$ vanishes.

\subsection{Non-Ideal Switching Transition}

In practice, the switch's finite output resistance and parasitic capacitance result in a time constant as:
\begin{equation}
\tau=R_{out}\cdot C_{out}
\end{equation}
which slows the transition between the supply rails. Hence, time mismatch can be approximated by a combination of three waveform segments:
1) constant voltage period, 2) ramp transition, and 3) exponential transition.
The Fourier coefficient of the total waveform is obtained by summing the Fourier coefficients of these segments.

\begin{equation}
C_1=C_{pulse}+C_{rmp}+C_{exp},
\end{equation}
and the output power is calculated as in equation \eqref{eq_pout}.

\subsection{Fourier Coefficient of the Ramp Section}

Consider a ramp waveform with voltage change $\Delta V$ occurring during a time interval $\Delta t$. The ramp can be expressed as:

\begin{equation}
v(t)=\frac{\Delta V}{\Delta t}t,
\end{equation}
for $0<t<\Delta t$.

Substituting this into the Fourier integral yields:

\begin{equation}
C_{rmp}=
\frac{\Delta V}{\Delta t\,T}
\frac{e^{-j\omega\Delta t}(j\omega\Delta t+1)-1}{\omega^2}.
\end{equation}

This expression represents the contribution of the ramp transition to the fundamental component.

\subsection{Fourier Coefficient of the Exponential Section}

When the output node charges/discharges through an RC network, while both PMOS and NMOS are in the ON state, the voltage follows an exponential trajectory:

\begin{equation}
v(t)=\Delta V\left(1-e^{-t/\tau}\right).
\end{equation}

The Fourier coefficient of this exponential segment is obtained by substituting the waveform into the Fourier integral

\begin{equation}
C_k=\frac{1}{T}\int_0^{\Delta t} v(t)e^{-jk\omega t}dt,
\end{equation}
Carrying out the integration results in:

\begin{equation}
C_{exp} = \frac{\Delta V}{T}
\left[\frac{1-e^{-j\omega\Delta t}}{j\omega}-\frac{1-e^{-\Delta t (1/\tau+j\omega)}}{1/\tau+j\omega}\right].
\end{equation}

This coefficient captures the spectral contribution of the RC charging/discharging behavior of the output node.

\printbibliography
\end{document}